\documentclass[sn-aps]{sn-jnl}

\jyear{2021}%

\theoremstyle{thmstyleone}%
\theoremstyle{thmstyletwo}%

\theoremstyle{thmstylethree}%

\def\be{\begin{equation}}
\def\ee{\end{equation}}
\def\bea{\begin{eqnarray}}
\def\eea{\end{eqnarray}}

\def\bi{\begin{itemize}}
\def\ei{\end{itemize}}

\usepackage[normalem]{ulem}

\usepackage{longtable,booktabs}
\usepackage{etoolbox}
\makeatletter
\patchcmd\longtable{\par}{\if@noskipsec\mbox{}\fi\par}{}{}
\makeatother
\IfFileExists{footnotehyper.sty}{\usepackage{footnotehyper}}{\usepackage{footnote}}
\makesavenoteenv{longtable}
\begin{document}

\title[Condensed Matter Physics in Big Discrete Time Crystals]{Condensed Matter Physics in Big Discrete Time Crystals}


\author*[1]{\fnm{Peter} \sur{Hannaford}}\email{phannaford@swin.edu.au, ORCID~Number: 0000-0001-9896-7284}

\author[2]{\fnm{Krzysztof} \sur{Sacha}}\email{krzysztof.sacha@uj.edu.pl, ORCID~Number: 0000-0001-6463-0659}

\affil*[1]{\orgdiv{Optical Sciences Centre}, \orgname{Swinburne University of Technology}, \orgaddress{\street{Hawthorn}, \city{Victoria 3122}, \country{Australia}}}

\affil[2]{\orgdiv{Institute of Theoretical Physics}, \orgname{Jagiellonian University}, \orgaddress{\street{ulica Profesora Stanisława Łojasiewicza 11}, \city{Krak\'ow}, \postcode{PL-30-348}, \country{Poland}}}


\abstract{
We review the application of discrete time crystals created in a Bose-Einstein condensate (BEC) of ultracold atoms bouncing resonantly on an oscillating atom mirror to the investigation of condensed matter phenomena in the time dimension.  Such a bouncing BEC system can exhibit dramatic breaking of time-translation symmetry, allowing the creation of discrete time crystals having up to about 100 temporal lattice sites and suitable for hosting a broad range of temporal condensed matter phenomena. We first consider single-particle condensed matter phenomena in the time dimension which include Anderson localization due to temporal disorder, topological time crystals, and quasi-crystal structures in time. We then discuss many-body temporal condensed matter phenomena including Mott insulator phases in time, many-body localization in time, many-body topological time crystals and time crystals having long-range exotic interactions. We also discuss the construction of two (or three) dimensional time lattices, involving the bouncing of a BEC between two (or three) orthogonal oscillating mirrors and between two oscillating mirrors oriented at 45-degrees. The latter configuration supports a versatile Möbius strip geometry which can host a variety of two-dimensional time lattices including a honeycomb time lattice and a Lieb square time lattice. Finally, we discuss the construction of a six-dimensional time-space lattice based on periodically driven BECs trapped in a three-dimensional optical lattice. 
}




\maketitle

\section{Introduction}\label{sec1}

In 2012 Frank Wilczek speculated whether crystalline structures
in space could be extended to the fourth dimension -- time \cite{Wilczek2012}? To create
such a `quantum time crystal', he considered the possibility of spontaneous
breaking of continuous time-translation symmetry in the ground state of a many-body system, in analogy with the
formation of an ordinary crystal in space. 
While a classical particle can perform periodic motion even if its energy is the lowest possible and a classical time crystal can form \cite{Shapere2012,Das2018,Shapere2019}, the quantum time crystal proposed by Wilczek turned out not to be feasible \cite{Bruno2013b,Watanabe2015,Watanabe2020}. However, his idea inspired further proposals in
time-independent \cite{Kozin2019} and time-dependent quantum systems. In particular, it
was shown that a \emph{periodically driven} many-body system can
spontaneously break discrete time-translation symmetry and
self-reorganize its motion so that it evolves with a period longer than
the drive \cite{Sacha2015}. To create such a `discrete time crystal', a range of
experimental platforms have been proposed, including a periodically
driven Bose-Einstein condensate \cite{Sacha2015}, driven spin-based systems \cite{Khemani16,ElseFTC,Yao2017,Russomanno2017,Pizzi2019a},
and driven open dissipative systems \cite{Gong2017,Sun2019,Cosme2019}. A number of experiments
providing evidence of discrete time crystals have been reported \cite{Zhang2017,Choi2017,Pal2018,Rovny2018,Kreil2018,Smits2018,Autti2018,
Kessler2020,Taheri2020,Autti2021,Kyprianidis2021,
Randall2021,Mi2021}. Several review articles and a monograph on time
crystals have been published \cite{Sacha2017rev,else2019discrete,khemani2019brief,
Guo2020,SachaTC2020,GuoBook2021}.

In this article we consider the case of a Bose-Einstein condensate (BEC)
of ultracold atoms bouncing resonantly on an oscillating atom mirror
such that the period of the bouncing atoms is equal to an integer
multiple $s$ of the period of the driving mirror \cite{Sacha2015}. Such a
system has been shown to exhibit dramatic breaking of discrete time-translation
symmetry in which the bouncing atoms can evolve with periods up to about
100 times longer than the period of the drive \cite{Giergiel2018a,Giergiel2020} (see also \cite{Surace2018,Pizzi2019a,Pizzi2021}). This allows the creation of big discrete time crystals
possessing a large number of temporal lattice sites, which are suitable
for realizing a broad range of condensed matter phenomena in the time
dimension \cite{Sacha15a,sacha16,delande17,Matus2021,
Mierzejewski2017,Giergiel2018b,Giergiel2018,
Giergiel2018c,Kosior2018,Zlabys2021,Kuros2021,Giergiel2021}. It has been shown for the case $s=2$ that such discrete time crystals are
robust against external perturbations \cite{Kuros2020} and quantum fluctuations \cite{Kuros2020,Wang2020} and live for extremely long times \cite{Wang2020,Wang2021}. Condensed
matter phenomena can also be investigated in photonic time crystals \cite{Lustig2018,Sharabi2021} and phase space crystals \cite{Guo2013,Guo2016,Guo2016a,Liang2017,Guo2020,
Guo2021,GuoBook2021}.

A resonantly driven bouncing BEC system allows effective temporal
lattice potentials to be engineered almost at will, by choosing suitable
Fourier components in the periodic driving function of the atom mirror.
This enables the geometry of the time lattice to be readily varied and
effective lattice potentials to be constructed for a broad
range of condensed matter phenomena in the time dimension. Use
of a resonantly driven many-body BEC also allows one to precisely
control the effective interparticle interaction and to engineer exotic
long‑range interactions in a time lattice by modulating the
interparticle s‑wave scattering length via a 
Feshbach resonance \cite{Giergiel2018}.

Temporal condensed matter phenomena that have been predicted to date
include Mott insulator phases in the time dimension \cite{Sacha15a}; Anderson
localization \cite{Sacha15a,sacha16,delande17,Giergiel2018a,
Sharabi2021,Matus2021} and many-body localization in the time
dimension \cite{Mierzejewski2017}; topologically protected edge states in time \cite{Lustig2018,Giergiel2018b}; quasi-crystal structures in time \cite{Giergiel2018,Giergiel2018c};
two-dimensional time lattices supporting a M\"obius strip geometry and
flat-band physics \cite{Giergiel2021}; and time-space crystals exhibiting a
six-dimensional quantum Hall effect \cite{Zlabys2021}.

In Sec.~\ref{b_tc}, we describe the physical principles underlying the
creation of big discrete time crystals in a periodically driven bouncing
BEC system and a proposed experiment for realizing big discrete time
crystals suitable for investigating condensed matter physics in the time dimension. In Sec.~\ref{s-p_condmat}, we discuss
single-particle condensed matter phenomena in the time dimension; in
Sec.~\ref{m-b_condmat} we describe many-body temporal condensed matter phenomena; and
in Sec.~\ref{md_timecrystals} we describe temporal condensed matter phenomena in
multi-dimensional time lattices.

\section{Big Discrete Time Crystals}
\label{b_tc}

{\it Single-particle system}

We first consider a single particle bouncing in the vertical direction $z$ on a hard-wall potential mirror which is oscillating with
frequency $\omega$ in the presence of strong transverse confinement.
Introducing gravitational units of length, energy and time, $l_{0}=(\hbar^{2}/m^{2}g)^{1/3}$, $E_{0} = mgl_{0}$, $t_{0}=\left(\hbar/mgl_{0}\right)^{1/3}$, and
transforming to the frame moving with the oscillating mirror, the
Hamiltonian of the system can be expressed as \cite{SachaTC2020}
\be
H = \ H_{0}(z,p) + H_{1}(z)f(t),
\label{h_init}
\ee
with the constraint $z\ge 0$, where $H_{0} =p^{2}/2 + z$, $H_{1} = z$ and
$f\left( t + 2\pi/\omega \right) = f\left( t \right)$ which
describes how the mirror oscillates in time. $p$ and \emph{m} are the
particle's momentum and mass and $g$ is the gravitational
acceleration.

Here, we are interested in the case of resonant driving, i.e., when the driving
frequency $\omega$ of the mirror is an integer multiple $s$ of the
bouncing frequency $\mathrm{\Omega}$ of the particle. It is convenient
to switch from the Cartesian position-momentum variables, $z$ and
$p$, to the action-angle variables, $I$ and $\theta\ $ \cite{Lichtenberg1992},
where the unperturbed Hamiltonian depends only on the new momentum (action
$I$), $H_{0}\left( I \right)$. Then, in the absence of the
driving, the solutions of the classical equations of motion are very
simple because the action is a constant of motion,
$I\left( t \right) =\text{constant}$, and the conjugate position variable
(angle $\theta$) evolves linearly in time,
$\theta\left( t \right) = \Omega\left( I \right)t + \theta\left( 0 \right)$,
where
$\Omega\left( I \right) = dH_{0}( I)/dI$. 

If a periodically evolving particle is resonantly driven,
i.e., if there is another part of the Hamiltonian,
$H_{1}\left( z \right)f\left( t \right)$, where
$f\left( t + 2\pi/\omega \right) = f\left( t \right)$ and
$s = \omega/\Omega\left( I_{s} \right)$ is an integer number,
then the classical effective time-independent Hamiltonian describing the motion of the
particle in the vicinity of the resonant trajectory has the form \cite{Giergiel2018,SachaTC2020} 
\be
H_{eff}=\frac{P^2}{2m_{eff}}+V_{eff}(\Theta), \quad V_{eff}(\Theta)=\sum_m h_{ms}(I_s)f_{-m}e^{ims\Theta},
\label{singheff}
\ee
where $P = I - I_{s}$, $I_{s}$ is the value of the action that
fulfills the $s:1$ resonance condition, and $m_{eff}^{-1}=d^2H_0(I_s)/dI_s^2$
is the inverse of the effective mass of the particle. The effective
potential in Eq.~(\ref{singheff}) depends on the components of the expansion of
$H_{1}$ in the action-angle variables,
$H_{1} = \sum_{n}^{}h_{n}\left( I \right)e^{in\Theta}$, and on the
Fourier components of
$f\left( t \right) = \sum_{k}^{}f_{k}e^{ik\omega t}$. The Hamiltonian
(\ref{singheff}) is time-independent because it is obtained in the frame moving along
the resonant trajectory
\be
\Theta=\theta-\frac{\omega t}{s}. 
\label{singTheta}
\ee

For pure harmonic driving,
$f\left( t \right) = \lambda\cos\left( \omega t \right)$, the
effective Hamiltonian (\ref{singheff}) can be expressed as \cite{Giergiel2018a} 
\be
H_{eff} = \frac{P^2}{2m_{eff}}+h_{s}(I_s)\cos(s\Theta).
\label{singheff1}
\ee
Plots of the action $I$ (or $P = I - I_{s}$) versus angle
$\Theta$ for a given resonance number $s$ lead to $s$ resonance islands in
phase space, which for a hard-wall potential mirror 
are stable with minimal chaotic motion for a driving strength
$\lambda\lesssim 0.2$, independent of \emph{s} and the oscillation
frequency $\omega$ \cite{Giergiel2020}. Note that the amplitude of the mirror oscillations
in the laboratory frame is $\lambda/\omega^2$, not
$\lambda$, because in order to obtain (\ref{h_init}) we have switched from the
laboratory frame to the frame oscillating with the mirror.

Switching to the quantum description (by quantizing the classical effective
Hamiltonian (\ref{singheff1}) or by performing a fully quantum effective description
of the resonant driving \cite{SachaTC2020,Giergiel2018,Berman1977}), eigenenergies of
$H_{eff}$ form energy bands $E_{n}\left( \kappa \right)$,
where $\kappa$ is a temporal analog of the quasi-momentum, and the corresponding
eigenstates have the form of Bloch waves
$e^{i\kappa\Theta}u_{n,\kappa}\left( \Theta \right)$, with
$u_{n,\kappa}\left( \Theta + 2\pi/s \right) = u_{n,\kappa}\left( \Theta \right)$.
Such crystalline behavior, which we deal with in the moving frame, will
be observed in the time domain when we return to the laboratory frame.
This is due to the fact that the transformation between the laboratory
and moving frames is linear in time \cite{sacha16,SachaTC2020,Giergiel2018}, cf. Eq.~(\ref{singTheta}). This
implies that if we fix the position in the laboratory frame close to the
resonant trajectory (i.e., $\theta = \text{constant}$ and
$I \approx I_{s}$), then the crystalline behavior that is observed
in the moving frame versus $\Theta$ will be observed as a function of
time in the laboratory frame because the Bloch waves take the form 
$e^{i\kappa\left( \theta - \omega t/s \right)}u_{n,\kappa}\left( \theta - \omega t/s \right)$.

If we are interested in the first energy band of such a crystalline
structure, the description of the particle can be reduced to the
tight-binding model \cite{Dutta2015}. That is, there are $s$ Wannier-like states
$w_{j}\left( z,t \right)$ corresponding to the first energy band which
are localized wave-packets evolving along the resonant trajectory with
period $sT,\ $where $T = 2\pi/\omega$, i.e.,
$w_{j}\left( z,t + sT \right) = w_{j}\left( z,t \right)$ and
$w_{j}\left( z,t + T \right) = w_{j + 1}\left( z,t \right)$ \cite{Sacha15a}.
The energy of the particle in this subspace (more precisely the
quasi-energy of the periodically driven particle \cite{Shirley1965}) takes the
form
\be
H_{eff}\approx -\frac{J}{2}\sum_{j=1}^s\left(a_{j+1}^*a_j+c.c.\right),
\label{singt-b}
\ee
where we have assumed that the wavefunction of the particle
$\psi\left( z,t \right) = \sum_{j = 1}^{s}a_{j}w_{j}\left( z,t \right)$.
In Eq.~(\ref{singt-b}), $J$ is the tunneling (or hopping) amplitude of the
particle between different localized wave-packets $w_{j}$.
\newline

\noindent
{\it Many-body system}

The second quantization formalism allows us to switch from a
single particle to many indistinguishable particles by introducing the
bosonic field operator $\hat{\psi}$, which can be expanded in
terms of operators that annihilate a boson in time-independent
single-particle basis states \cite{Pethick2002}. In the case of resonantly driven many-body
systems, such an approach can be reformulated in terms of a
time-periodic basis if we consider a many-body Floquet Hamiltonian and
extend the Hilbert space of the system by the temporal degree of freedom \cite{SachaTC2020}. In practice, this relies on expansion of the field operator
$\hat{\psi}$ in terms of operators $\hat a_{j}$ that annihilate a
particle in basis states $\phi_{j}\left( z,t \right)$ defined in the
extended Hilbert space. That is, the $\phi_{j}\left( z,t \right)$'s
form a complete basis for a particle moving in 1D space at any time
$t$ and are periodic with period $nT$, where
$T = 2\pi/\omega$ is the period of the external (mirror)
driving and $n$ can be any integer number not necessarily $n = 1$
--- in such a basis the spectrum of the Floquet Hamiltonian is periodic
with the period $\hbar\omega/n$ \cite{SachaTC2020}. Clearly, we cannot
afford a complete expansion of the field operator
$\hat{\psi}\left( z,t \right)$ and need to introduce some
truncation.

The description of a resonantly bouncing single particle, with the
Hamiltonian~(\ref{h_init}),
can be reduced to the tight-binding model, Eq.~(\ref{singt-b}), if we restrict the
Hilbert space to the resonant subspace spanned by $s$ localized
wave-packets $w_{j}\left( z,t \right)$ moving periodically along the
$s:1$ resonant trajectory. This subspace is related to the first
energy band of the single-particle effective Hamiltonian, Eq.~(\ref{singheff1}). If
the interaction energy per particle is much smaller than the energy gap
between the first and second energy bands of the Hamiltonian (\ref{singheff1}), then,
in the many-body case, we may restrict our analysis to the first energy
band, and the many-body Floquet Hamiltonian takes the form of a
Bose-Hubbard model in time \cite{SachaTC2020}
\bea
\hat H_F&=&\frac{1}{sT}\int\limits_0^{sT}dt \int dx\;\hat\psi^\dagger\left[H(t)-i\frac{\partial}{\partial t}+\frac{g_0}{2}\hat\psi^\dagger\hat\psi\right]\hat\psi 
\\ &\approx& -\frac{J}{2}\sum_{j=1}^s\left(\hat a_{j+1}^\dagger\hat a_j+H.c.\right)+\frac12 \sum_{i,j=1}^sU_{ij}\hat a_i^\dagger \hat a_j^\dagger \hat a_j \hat a_i,
\label{manybh}
\eea
where $g_{0}$ is the strength of the contact interaction between
particles which model the van der Waals interactions between ultracold
atoms \cite{Pethick2002}. In Eq.~(\ref{manybh}) we have assumed bosonic atoms and restricted
the Hilbert space to the resonant subspace; that is,
$\hat{\psi}\left( z,t \right) \approx \sum_{j = 1}^{s}w_{j}\left( z,t \right){\hat{a}}_{j}$,
where ${\hat{a}}_{j}$ are the standard bosonic annihilation
operators and $w_{j}$ are the Wannier-like localized wave-packets
which evolve along the resonant trajectory with period $sT$,
i.e., $w_{j}\left( z,t + sT \right) = w_{j}\left( z,t \right)$, cf.
Eq.~(\ref{singt-b}). The tunneling amplitude $J$ describes the hopping of atoms
between neighboring wave-packets in the time domain, i.e.,
between $w_{j}\left( z,t \right)$ and
$w_{j + 1}\left( z,t \right) = w_{j}\left( z,t + T \right)$. The
interaction coefficients $U_{ij}$ depend on the overlap of the
densities of the Wannier-like states, i.e.,
\be
U_{ij}=\frac{2-\delta_{ij}}{sT}\int\limits_0^{sT} dt \int dz\; g_0 \rvert w_i(z,t)\rvert^2 \;\rvert w_j(z,t)\rvert^2,
\label{manyuij}
\ee 
and in principle can describe effective long-range interactions in
the Bose-Hubbard model. However, in a typical situation, the effective
long-range interactions are negligible because they are orders of
magnitude weaker than the on-site interactions in Eq.~(\ref{manybh}) \cite{SachaTC2020}.

Within the mean-field approximation, we look for time-periodic $N$-body states in the form of a product
state $\phi\left( z_{1},t \right)\phi\left( z_{2},t \right)\ldots.\phi\left( z_{N},t \right)$.
The resonant subspace is spanned by \emph{s} single-particle Floquet
states or equivalently by \emph{s} localized wave-packets
$w_{j}\left( z,t \right)$. Thus, the mean-field periodic
solutions within the resonant subspace can be expanded in the
time-periodic basis
$\phi\left( z,t \right) = \sum_{j = 1}^{s}a_{j}w_{j}\left( z,t \right)\text{.\ }$The
coefficients $a_{j}$, which satisfy
$\sum_{j = 1}^{s}{{{\rvert a}_{j}\rvert}^{2} = 1,}$ correspond to extremal values
of the quasi-energy of the system per particle
\be
E_{F}\approx - \frac{J}{2}\sum_{i = 1}^{s}{(a_{i + 1}^{*}}a_{i} + c.c.) + \frac{N}{2}\sum_{i,j = 1}^{s}U_{ij}{{\rvert a}_{i}\rvert }^{2}{{\rvert a}_{j}\rvert}^{2},
\ee
which can be found by solving the corresponding Gross-Pitaevskii
equation \cite{Pethick2002}. The mean-field approach is very useful in the
description of the experiments on discrete time crystals in BECs and we will use it in the following.
\newline

\noindent
{\it Discrete time crystal: $s = 40$}

As an example, we consider a BEC of ultracold atoms bouncing resonantly
on an oscillating atom mirror for the case of the $40 : 1$ resonance
($s= 40$) with driving amplitude $\lambda = 0.2$ and driving
frequency $\omega = 4.9$ (both in gravitational units) \cite{Giergiel2018a}. When the
attractive interactions are sufficiently weak, i.e.,
$\rvert g_{1D}N\rvert \leq 1.6\times 10^{- 3}$ (where
$g_{1D} = g_{0}{(\omega}_{\bot}/2\pi) < 0$ and $\omega_{\bot}$ is
the trapping frequency of the transverse confinement) the lowest energy
state of the Bose-Hubbard Hamiltonian (\ref{manybh}) is well approximated by the mean-field
solution
$\psi\left( z_{1},\ldots..z_{N},t \right) = \prod_{i = 1}^{N}{\phi\left( z_{i},t \right)}$,
where 
\be
\phi\left( z,t \right) = \frac{1}{\sqrt{s}}\sum_{j = 1}^{s}w_{j}\left( z,t \right).
\label{GPsol}
\ee
Although the Wannier states $w_{j}\left( z,t \right)$ are periodic
with period \emph{sT}, the wavefunction $\phi\left( z,t \right)$ is
periodic with period \emph{T} because after each period \emph{T} the
wave-packets exchange their role, i.e.,
$w_{j + 1}\left( z,t \right) = w_{j}\left( z,t + T \right)\text{.\ }$Thus,
the mean-field time-periodic solution
$\psi\left( z_{1},\ldots..z_{N},t \right)$ preserves the discrete
time-translation symmetry of the many-body system.

However, when the attractive interactions are sufficiently strong,
i.e., $\rvert g_{1D}N\rvert > 1.6 \times 10^{- 3}$, the mean-field
Gross-Pitaevski solution (\ref{GPsol}) becomes dynamically unstable and \emph{s}
new stable periodic solutions are created which are a non-uniform
superposition of the Wannier wave-packets $w_{j}$ and therefore are evolving with a
period $s= 40$ times longer than the driving period \emph{T} and
consequently break the discrete time-translation symmetry of the
many-body Hamiltonian. For stronger interactions ${\rvert g}_{1D}N\rvert \gtrapprox 0.1$
the stable periodic solutions reduce practically to the single localized
wave-packets
$\phi\left( z,t \right)\approx w_{j}\left( z,t \right)$.

In order to recover the discrete time-translation symmetry of the
many-body system we can, in principle, prepare a superposition of the
mean-field product states
\be
\psi\left( z_{1},\ldots..z_{N} \right)\approx \frac{1}{\sqrt{s}}\sum_{j = 1}^{s}{\prod_{i = 1}^{N}{w_{j}\left( z_{i},t \right),}}
\label{SchCat}
\ee
which in the second quantization formalism reads
\be
\vert \psi\rangle\approx\frac{1}{\sqrt{s}}(\vert N,0,\ldots.,0\rangle + \vert 0,N,0\ldots.,0\rangle + \ldots + \vert 0,\ \ldots.,0,N\rangle.
\ee
The state (\ref{SchCat}) is a good approximation to the many-body Floquet state
corresponding to the lowest energy state of the Hamiltonian (7) for
${\vert g}_{1D}N\vert \gtrapprox 0.1$. It is a superposition of $s= 40$
Bose-Einstein condensates and because $\langle w_{j}\vert w_{j^{'}}\rangle=\delta_{jj^{'}}$ it is also a Schr\"odinger cat-like state. It
evolves with the driving period \emph{T} but after, for example,
measurement of the position of a particle, the Schr\"odinger cat state
collapses, the discrete time-translation symmetry is broken, and the
system starts evolving with a period $s = 40$ times longer than the
driving period \emph{T.}

\begin{figure}[t]
\centering
\includegraphics[width=0.44\textwidth]{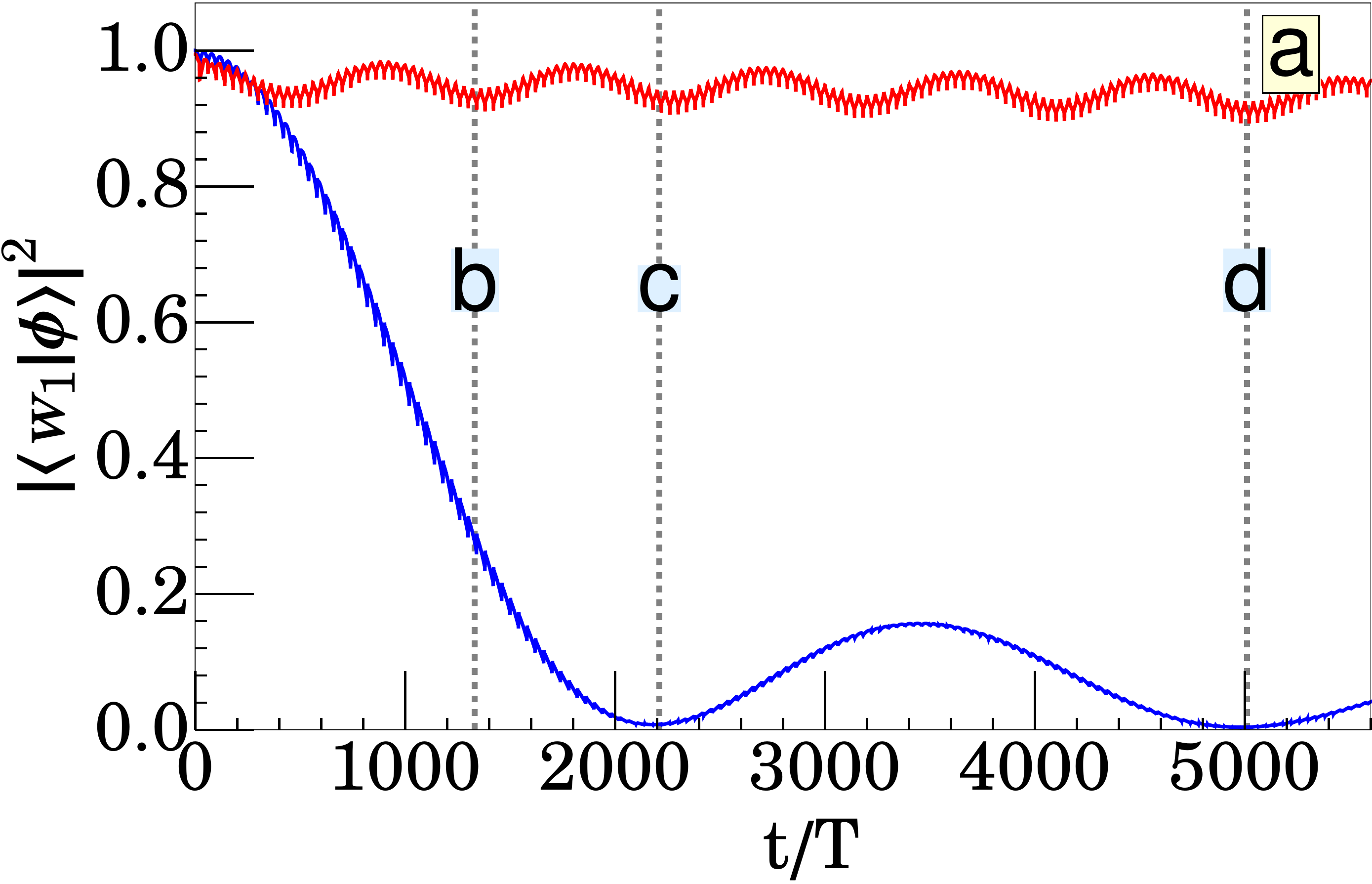}
\includegraphics[width=0.45\textwidth]{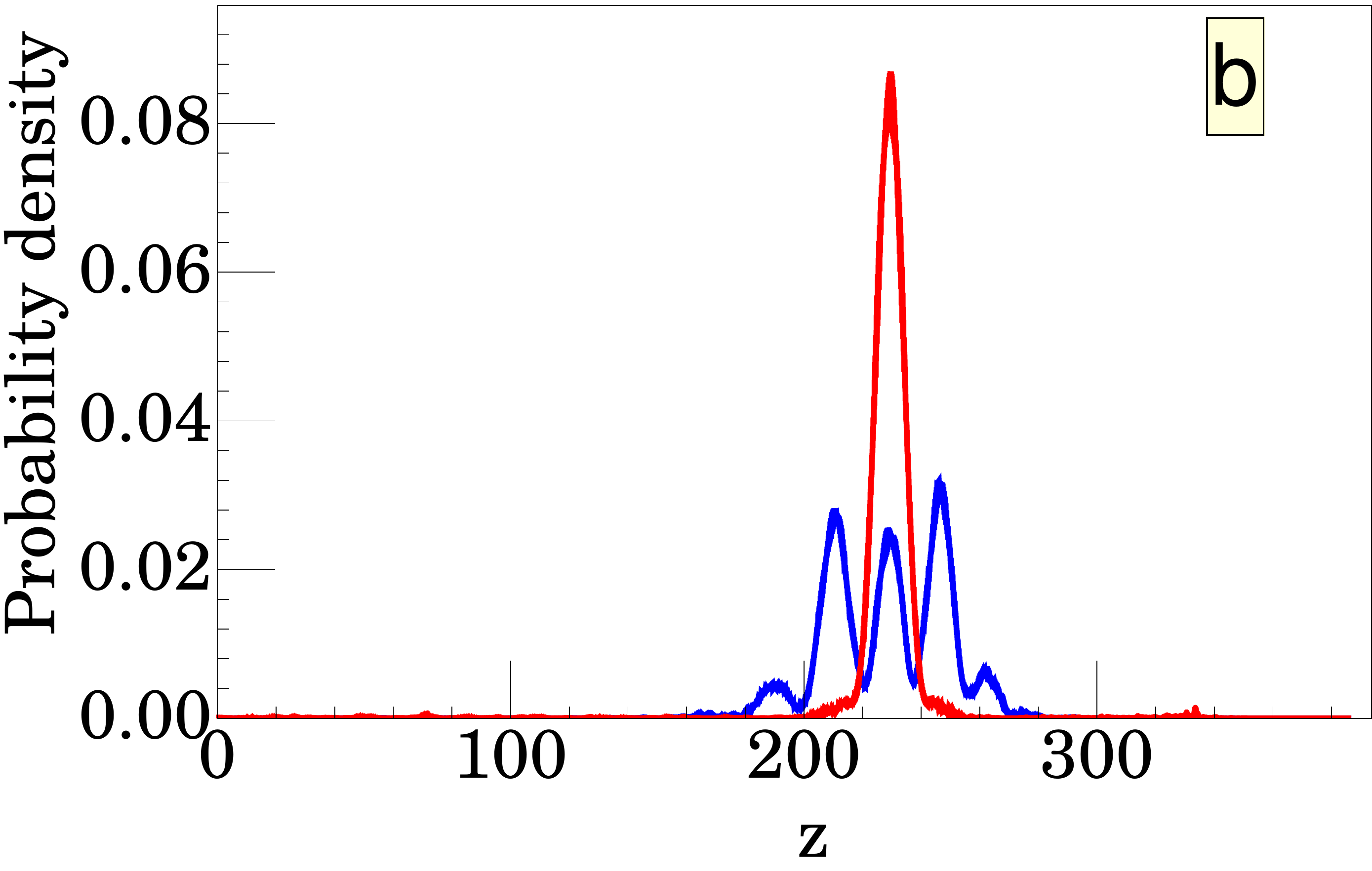}
\includegraphics[width=0.45\textwidth]{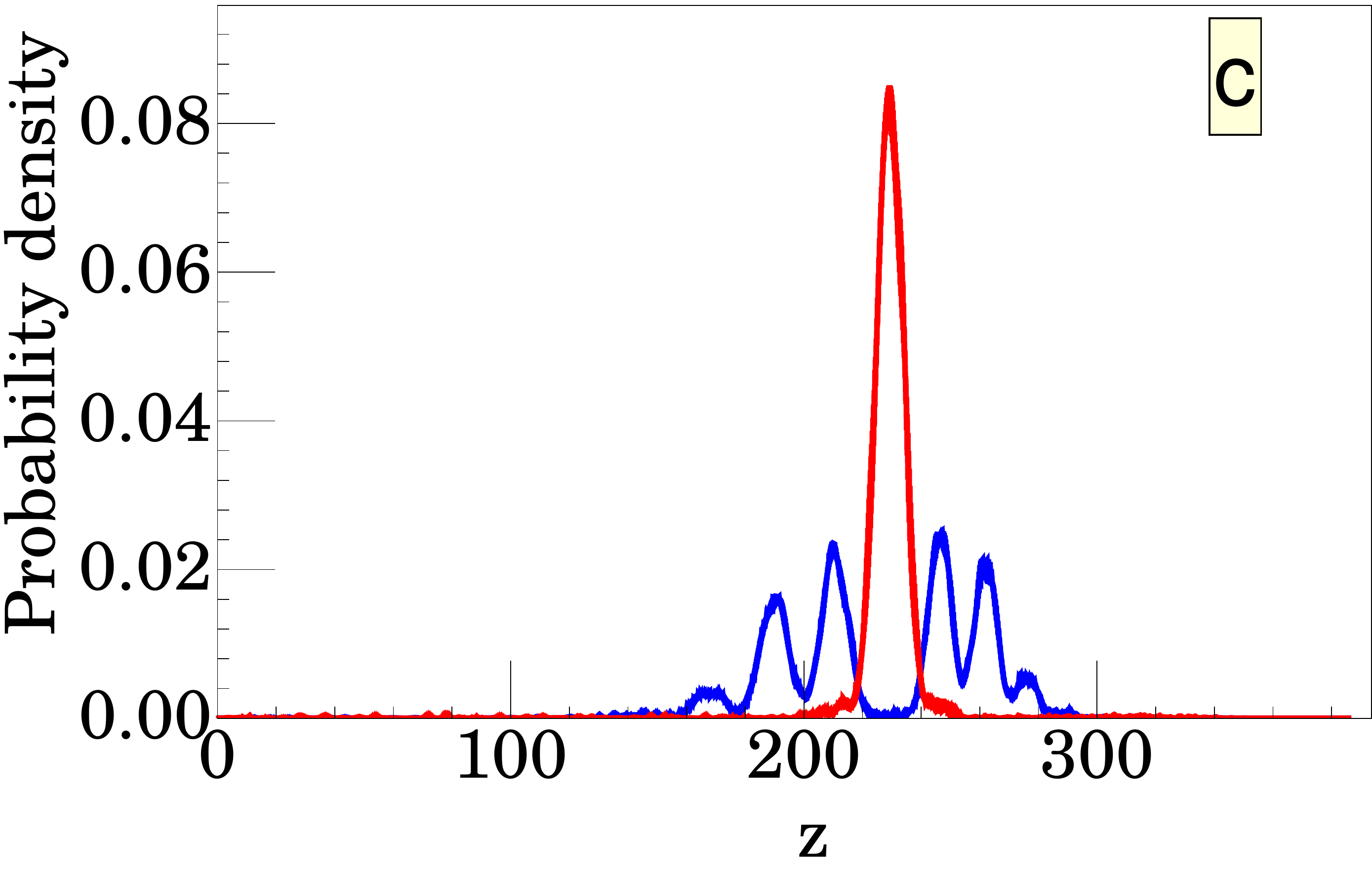}
\includegraphics[width=0.45\textwidth]{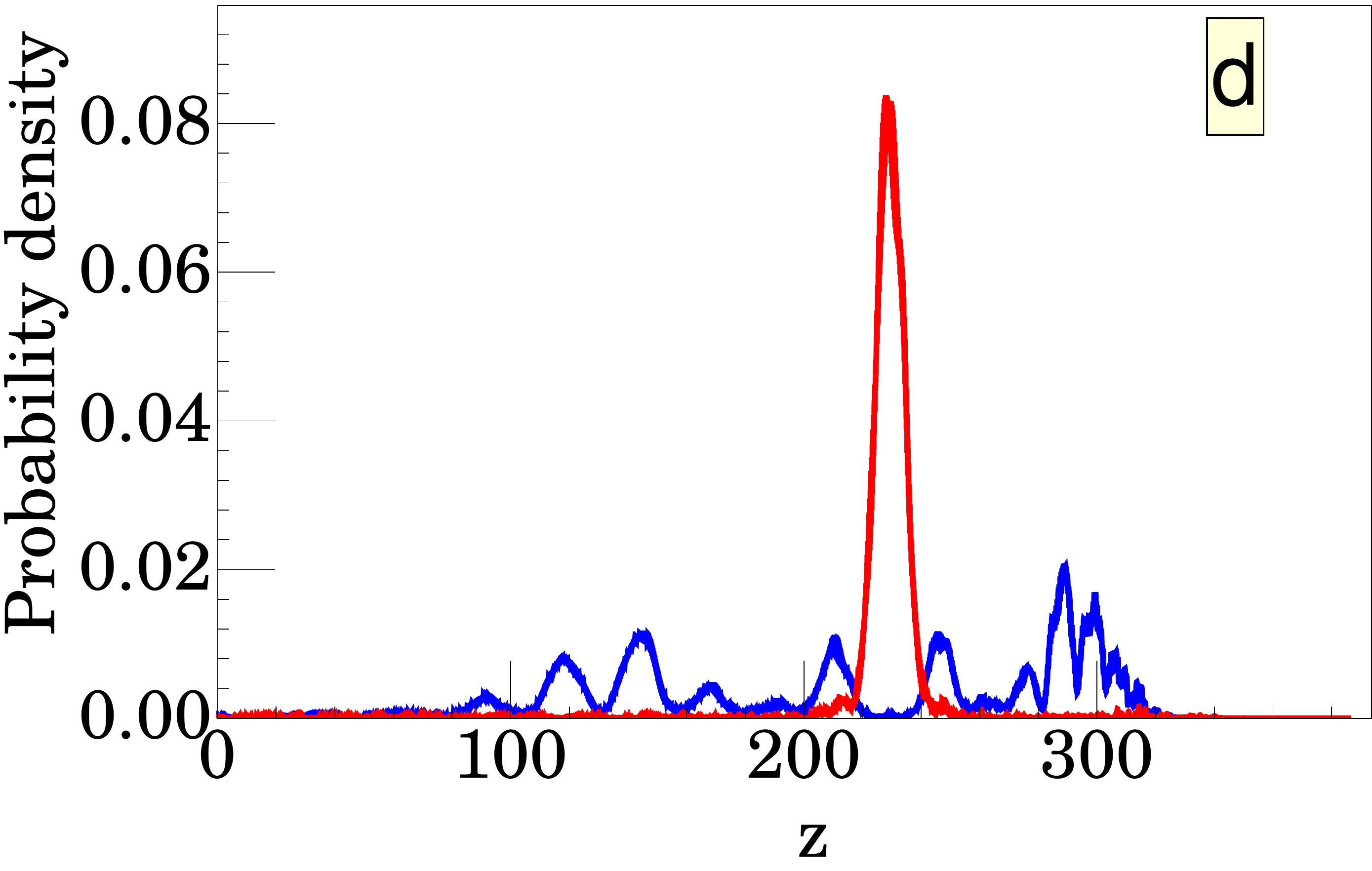}
\caption{Discrete time crystal formation for \emph{s} = 40. In the
absence of interactions (blue curves), atoms tunnel from the initial
state to neighboring Wannier wave-packets which is indicated by the
decrease of the overlap between the time evolving wavefunction $\phi(z,t)$ and the
initially chosen $w_1(z,0)$ Wannier state (blue curve in (a)). When a sufficiently
strong attractive interaction is turned on
($g_{1D}N = - 0.12)$, the system chooses a periodic solution
evolving with a period 40 times longer than the period expected from the
symmetry of the Hamiltonian, and the discrete time-translation symmetry
is broken (red curves). In (b)-(d), the densities of interacting
(noninteracting) atoms are presented with red (blue) curves at different
moments in time: (b) $t/T = 1330$, (c) 2210 and (d) 5010.
Parameters: $\lambda = 0.2$, $\omega = 4.9$. Reprinted from \cite{Giergiel2018a}.}
\label{DTC}
\end{figure}

In an experiment it will not be possible to prepare the Schr\"odinger cat
state (\ref{SchCat}). However, it should be straightforward to prepare a symmetry
broken state
\be
\psi\left( z_{1},\ldots..z_{N},0 \right) = \prod_{i = 1}^{N}{\phi\left( z_{i},0 \right)},
\ee
with $\phi\left( z,0 \right)\approx w_{j}\left( z,0 \right)\ $if
a BEC of ultracold atoms is prepared in a harmonic trap located at the
classical turning point above the oscillating atom mirror and the trap
is turned off at the moment the mirror is in its downward position \cite{Giergiel2018a}. Then the BEC starts evolving along the classical $40 : 1$
resonant orbit. If the attractive interactions are too weak, then atoms
start tunneling to neighboring wave-packets $w_{j - 1}$ and
$w_{j + 1}$ and after time
$t \approx 2.4/J\approx 2210T$
they totally leave the initial wave-packet $w_{j }$, as indicated by the decrease of the overlap between the time-evolving wavefunction $\phi(z, t)$ and the initially chosen Wannier wave-packet $w_1(z, 0)$ in Fig.~\ref{DTC}(a) (blue curves) and by the probability density, which is the experimental observable, in Fig.~\ref{DTC}(b)-(d) (blue curves). However, when the attractive interactions are sufficiently
strong (e.g., $g_{0}N =- 0.12$), the initial wavefunction
$\phi\left( z,0 \right)\approx w_{1}\left( z,0 \right)$
reproduces the symmetry broken state and no tunnelling to neighboring
wave-packets is observed, even for a time evolution as long as
5000\emph{T} (Fig.~\ref{DTC}, red curves), indicating the stability of the
discrete time crystal.
\newline

\noindent
{\it Gaussian-shape potential mirror}

In the above analysis we have assumed bouncing of particles from a hard-wall
potential mirror, for which the optimal driving strength to produce
stable resonance islands is
$\lambda\approx 0.2$  for $s \gg 1$ \cite{Giergiel2020}. In an experiment it
is convenient to use a repulsive light-sheet as the atom mirror, which
is well approximated by a Gaussian shape potential. For such a mirror,
the same time crystal phenomena as for a hard-wall potential mirror can
be realized, but the optimal driving strength needed to have the same
effect is typically an order of magnitude larger, which makes the
driving more accessible experimentally \cite{Giergiel2020}. In the case of a
light-sheet atom mirror, the driving can be conveniently implemented by
modulating the light intensity.
\newline

\noindent
{\it Robustness of discrete time crystals}

The robustness of a discrete time crystal created in a BEC bouncing
resonantly on a periodically driven mirror has been investigated
for the cases $s = 2$ and 4 \cite{Kuros2020}. The simulated effects of
imperfections in the driving and in the initial state are summarized in
the form of a phase diagram of the displacement $\epsilon$ of the initial state
from the classical turning point for perfect resonant
bouncing ($z = h_{0}$) versus the interaction strength $g_{0}N$ required to create a
discrete time crystal, with the
average atom momentum and the width of the initial wave-packet chosen
randomly \cite{Kuros2020}. For $s = 2$, the range of displacements around the
classical turning point is about
$\Delta\epsilon/h_{0}\approx 0.2$ for
$g_{0}N \approx - 0.02$, $\lambda = 0.12.$ For $s \gg 1$, the
range of displacements was estimated to be
$\Delta\epsilon/h_{0}\approx 8\sqrt{\lambda}/s\pi$
\cite{Kuros2020}, which is about
$0.03$ for $s = 40$,
$\lambda\approx0.2$. 
\newline

\noindent
{\it Many-body effects}

It has been observed that a BEC initially prepared in the ground state of an optical lattice potential or ultra-cold atoms in free space can be gradually depleted due to single-particle heating and two-particle processes when a time periodic perturbation is turned on \cite{Choudhury2014,Reitter2017,Li2019}. However, in the present article we consider a different situation to that where ultra-cold atoms are not initially prepared in an eigenstate of the time-independent Hamiltonian where the mirror does not move but the atoms are loaded to a resonant Hilbert  subspace of the periodically driven system. If, in the course of time evolution, the atoms do not leave this subspace, our predictions of the time crystal phenomena are valid. 

\begin{figure}[t]
\centering
\includegraphics[width=0.95\textwidth]{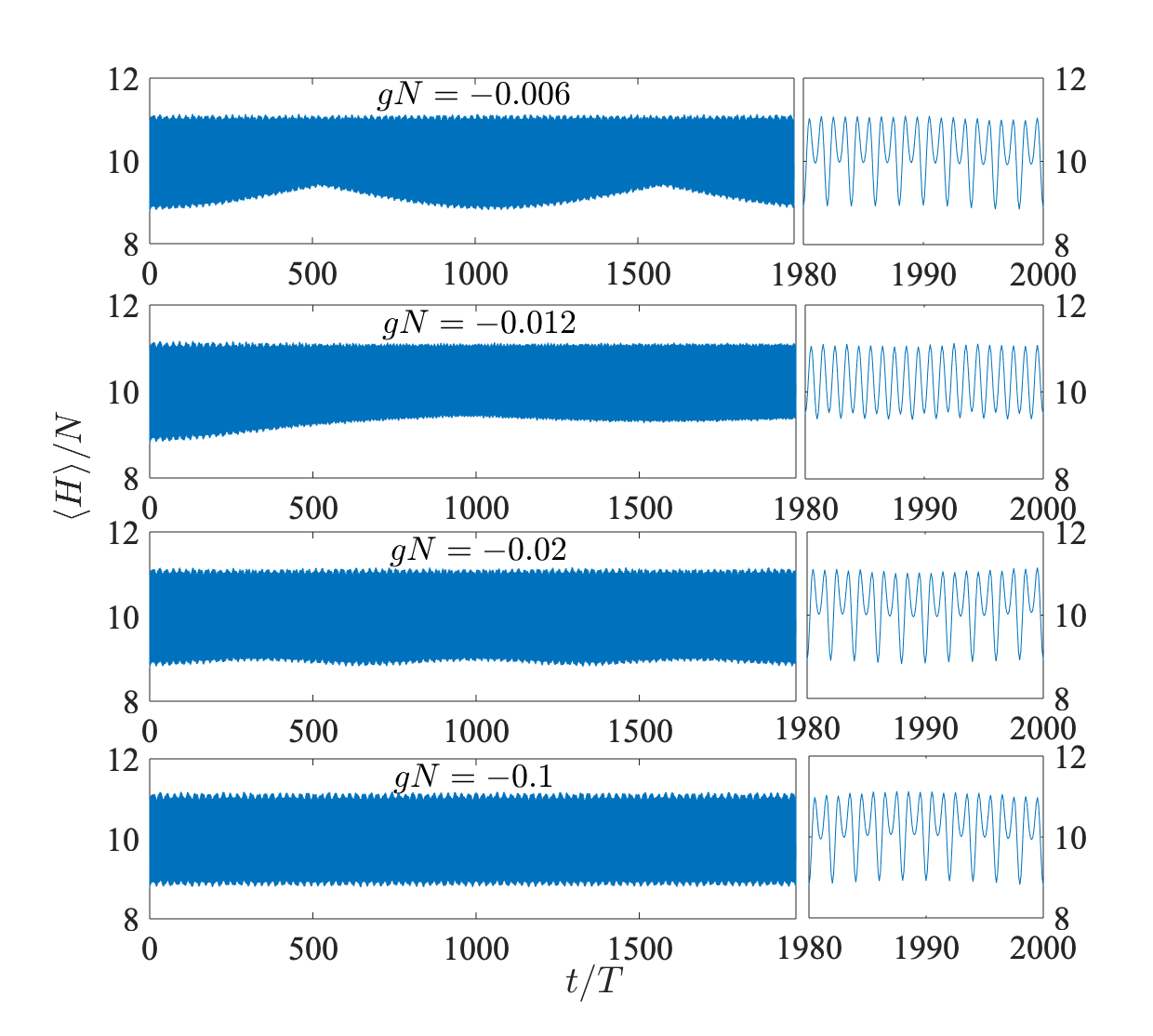}
\caption{Mean energy of the atoms (i.e., the sum of the kinetic energy, the potential energy in the gravitational field and the interaction energy) which are released from a harmonic trap and start bouncing on the oscillating mirror along the $2:1$ ($s=2$) resonant trajectory. The presented results are obtained within the TWA for $N=600$ \cite{Wang2020} and they correspond to the interaction strength $g_0N$ close to the critical value ($g_0N=-0.012$ \cite{Wang2021}) for the formation of a DTC and deep in the DTC regime ($g_0N=-0.1$). In all the cases, there is no signature of heating. Reprinted from \cite{Wang2020}.}
\label{heating_lack}
\end{figure}

So far, we have mainly used the mean-field approximation. The effects of quantum many-body phenomena and possible heating on the
formation of a discrete time crystal in a BEC bouncing resonantly on a
periodically driven mirror have been studied for \emph{s} = 2 using the
Bogoliubov approach \cite{Kuros2020} and also using a full multi-mode quantum treatment based on
the truncated-Wigner approximation (TWA) \cite{Wang2020}. The Bogoliubov and
TWA calculations agree broadly with the mean-field (single-mode)
calculations for times out to at least 2000 driving periods, except for
interaction strengths $g_{0}N$ very close to the critical value for
discrete time crystal formation where the TWA calculations predict a
quantum depletion up to about 260 atoms out of a total of 600 which results from atoms escaping to essentially just a
second mode (for $s=2$). The TWA calculations also predict that
the mean energy per particle does not increase significantly 
out to at least 2000 driving periods, indicating that the system reaches
a steady state with no net energy pumped from the drive, and thus there
is no evidence of thermalization on this time scale, see Fig.~\ref{heating_lack}. Additional many-body
calculations for $s= 2$ based on a two-mode model derived
from standard quantum field theory indicate that the discrete time
crystals survive for times out to at least 250,000 driving periods \cite{Wang2021}.
\newline

\noindent
{\it Size of discrete time crystals}

For a BEC bouncing resonantly on a periodically driven hard-wall
potential mirror, the drop height required to tune to a given resonance
\emph{s} can be expressed as \cite{Giergiel2020}
\be
h_{0} =(\pi/2)^{4/3} (\alpha/\lambda)^{1/3} s^{4/3},
\ee
where all quantities are in gravitational units and
$\alpha =2\pi^2s^2\lambda/\omega^6$ is a universal
parameter which we choose to be $\alpha = 0.456$ so that the energy gap between
the first and second energy bands of the quantum version of the effective Hamiltonian
(\ref{singheff1}) is $\Delta E/J\approx 10$ for $\lambda =0.2$ \cite{Giergiel2020}. This
$\Delta E/J$ value is sufficently large to ensure
validity of the single-band Hamiltonian (\ref{manybh}) while sufficiently small to
allow all atoms to tunnel to neighbouring wave-packets during a reasonably small number of bounces ($N_b = 54$) in the absence of interactions.
The driving frequency $\omega$ required for a given resonance $s$ and drop height
\emph{h}\textsubscript{0} is
\be
\omega = \pi s/\sqrt{2h_0}.
\label{omega_h0}
\ee

\begin{table}[h]
\begin{center}
\begin{minipage}{\textwidth}
\caption{Dependence of parameters for the case of a \textsuperscript{39}K atom on the
size of a discrete time crystal 
for a hard-wall potential mirror and $\lambda= 0.2$,
$\Delta E/J=10$, $N_b = 54$.
Gravitational units for \textsuperscript{39}K are
$l_0 = 0.647 \mu$m, $t_0 = 0.256$ms. Adapted from \cite{Giergiel2020}.}
\label{Table1}
\begin{tabular*}{\textwidth}{@{\extracolsep{\fill}}lllll@{\extracolsep{\fill}}}
\toprule
$s$ & $\lambda l_0/\omega^2$  & $h_0l_0$ & $\omega/(2\pi t_0)$ & $t_{tunnel}t_0$ \\
 & [nm] & [$\mu$m] & [kHz] & [s] \\
\midrule
10 & 13.6 & 34 & 1.92 & 0.28\tabularnewline
20 & 8.56 & 85 & 2.42 & 0.45\tabularnewline
30 & 6.53 & 145 & 2.77 & 0.59\tabularnewline
40 & 5.38 & 213 & 3.05 & 0.71\tabularnewline
50 & 4.64 & 281 & 3.28 & 0.83\tabularnewline
60 & 4.11 & 365 & 3.49 & 0.93\tabularnewline
70 & 3.71 & 449 & 3.67 & 1.03\tabularnewline
80 & 3.40 & 536 & 3.84 & 1.13\tabularnewline
90 & 3.14 & 627 & 3.99 & 1.22\tabularnewline
100 & 2.93 & 722 & 4.13 & 1.31\tabularnewline
\botrule
\end{tabular*}
\end{minipage}
\end{center}
\end{table}

Table~\ref{Table1} summarizes calculated parameters for different size discrete time crystals in
the range $s=10-100$ for the case of a \textsuperscript{39}K atom and a hard-wall potential mirror with $\lambda= 0.2$,
$\Delta E/J = 10$, $N_b = 54$ \cite{Giergiel2020}.
In a time crystal experiment, we require a relatively large drop height ($h_0l_0$) to allow high spatial
resolution probing of the atom density over a range of distances between the classical turning point and the oscillating mirror,
a relatively large mirror oscillation amplitude ($\lambda l_0/\omega^2$ in the laboratory
frame), and a relatively short tunnelling time ($t_{tunnel}t_0$) to allow the
experiment to be performed in times shorter than the lifetime of the
bouncing BEC. For $s = 10$, the drop height is only 34~$\mu$m which limits the 
spatial resolution for probing over a range of distances $z$. For $s =100$, the amplitude of the drive is only about 2.9~nm for the case
of a hard-wall potential mirror, but the amplitude will be about an
order of magnitude larger for a realistic soft Gaussian potential
mirror.
\newline

\noindent
{\it Experimental system}

A suitable atomic system for realizing a discrete time crystal in a BEC
bouncing resonantly on an oscillating mirror is bosonic potassium-39.
Potassium-39 has a broad Feshbach resonance
centred at 402~G \cite{DErrico2007} which allows one to magnetically tune the inter-atomic s-wave
scattering length, and hence the interaction 
strength,  in the region near $a_{s} = 0$ with
very high precision, i.e.,
$\rvert da_{s}/dB\rvert_{a_{s} = 0} \approx\;\rvert a_{bg}/\Delta\rvert = 0.56a_0/$G
(where $a_{bg}$ is the background scattering length, $\Delta$
is the width of the Feshbach resonance, and \emph{a}\textsubscript{0} is
the Bohr radius). Potassium-39 also has resonance lines at a convenient
wavelength for diode laser sources: 767 nm (D2) and 770 nm (D1).

In a proposed experiment currently being set up in Melbourne, a \textsuperscript{39}K BEC is prepared in a
crossed optical dipole trap \cite{Salomon2014} in the vertical plane, and
positioned at height $z =h_0$ corresponding
to the selected $s:1$ resonance above an oscillating light-sheet
atom mirror. The BEC is then released from the optical dipole trap to
fall on to the oscillating light-sheet mirror under strong transverse
confinement. The light-sheet atom mirror is driven with frequency $\omega$
corresponding to the drop height $h_0$, Eq.~\eqref{omega_h0}, and the
amplitude $\lambda$ of the drive is adjusted to produce stable resonance islands
located around periodic orbits with period $sT$ \cite{Giergiel2018a,Giergiel2020}.

The atom density is recorded at fixed positions between the classical
turning point and the oscillating atom mirror and at different moments
in time for different particle interaction strengths $g_{1D}N$ by varying the interparticle
scattering length $a_{s}$ via the broad Feshbach resonance at 402 G \cite{DErrico2007}.
The time for atoms to tunnel from the initial wave-packet to neighboring
wave-packets (which is the time scale of the system dynamics in a time
lattice) in the absence of interactions is about 0.7 s for $s=40$
\cite{Giergiel2020}. When the particle interaction strength $\rvert g_{1D}N\rvert$ is
raised above a critical value to spontaneously break the
time-translation symmetry, a stable localized wave-packet evolves
without tunneling to other wave-packets, i.e., a discrete time crystal
is formed (Table~\ref{Table1}), see Fig.~\ref{DTC} (red curves).

\section{Single-Particle Condensed Matter Phenomena in the Time Dimension}
\label{s-p_condmat}

Spontaneous breaking of continuous {\it space} translation symmetry into discrete space translation symmetry results in the formation of ordinary space crystals. However, crystalline structures in space can also be created externally without any spontaneous process. For example, by imposing periodic behavior of the refractive index in space in dielectric materials, one obtains photonic crystals in which the propagation of electromagnetic waves possesses similar properties to the transport of an electron in a space crystal \cite{Joannopoulos_Book}. Another example is ultracold atoms in optical lattice potentials which are routinely created in the laboratory by means of electromagnetic standing waves \cite{Lewenstein2007Adv}. There, ultracold atoms behave like solid state systems but the crystalline structures in space do not emerge spontaneously. 

Crystalline structures can also be created by means of a proper periodic driving of the system. It has been shown by Guo and coworkers that a resonantly driven harmonic oscillator reveals a crystalline structure but in phase space \cite{Guo2013,Guo2016,Guo2016a,Liang2017,Guo2020,
Guo2021,GuoBook2021}. Consider a periodically kicked harmonic oscillator where an additional perturbation is turned on periodically during very short periods of time, $H=(p^2+x^2)/2+KT\cos x\sum_n\delta(t-nT)$, where the driving period $T$ is $s$ times shorter than the harmonic oscillator period ($s$ is integer) and $K$ is the strength of the kicking. Applying the rotating wave approximation, one obtains an effective time-independent Hamiltonian that describes different crystalline structures in the 2D phase space for different $s:1$ resonant drivings. The $x$ and $p$ operators do not commute, and to plot the crystal structures one has to use, for example, the coherent state representation \cite{Liang2017}. For $s=4$, a square lattice is observed in phase space; for $s=3$ and $s=6$, a hexagonal lattice structure is observed; while for $s=5$ and $s\ge 7$, a quasi-crystal structure is observed. The non-commutative geometry in phase space can lead to topological properties of the system. For $s=4$, the effective Hamiltonian has a simple form, $H_{eff}=K(\cos x+\cos p)/2$, and its energy spectrum depends on the value of the effective Planck constant, $\hbar_{eff}$, which can be controlled experimentally \cite{Liang2017}. If $\hbar_{eff}$ is a rational number, then there exists an abelian group related to translations in phase space which is a symmetry group of the system. Energy levels of $H_{eff}$ form topologically nontrivial bands which, similar to the case of the quantum Hall effect, reveal a characteristic Hofstadter butterfly structure as a function of different rational values of $\hbar_{eff}$ \cite{Liang2017}.

In the present article we focus on periodically driven systems which reveal crystalline structures but in the time domain. In the case of space crystals one is interested in a regular distribution of particles in space which can be observed at a fixed moment of time, i.e., at the moment of experimental detection. Switching from space to time crystals, we have to exchange the roles of space and time \cite{sacha16}. We fix the position in space, i.e., we locate a detector at a certain space point, and ask if the probability of clicking of the detector is periodic in time. Obviously, the periodic evolution of many different systems fulfills this criterion but we are interested in systems whose periodic behavior in time is described by condensed matter models \cite{Sacha2017rev,SachaTC2020}. 

In Sec.~\ref{b_tc} we introduced the general approach for the realization of condensed matter in the time domain and now we are ready to present different examples. 
\newline

\begin{figure}[t]
\centering
\includegraphics[width=0.45\textwidth]{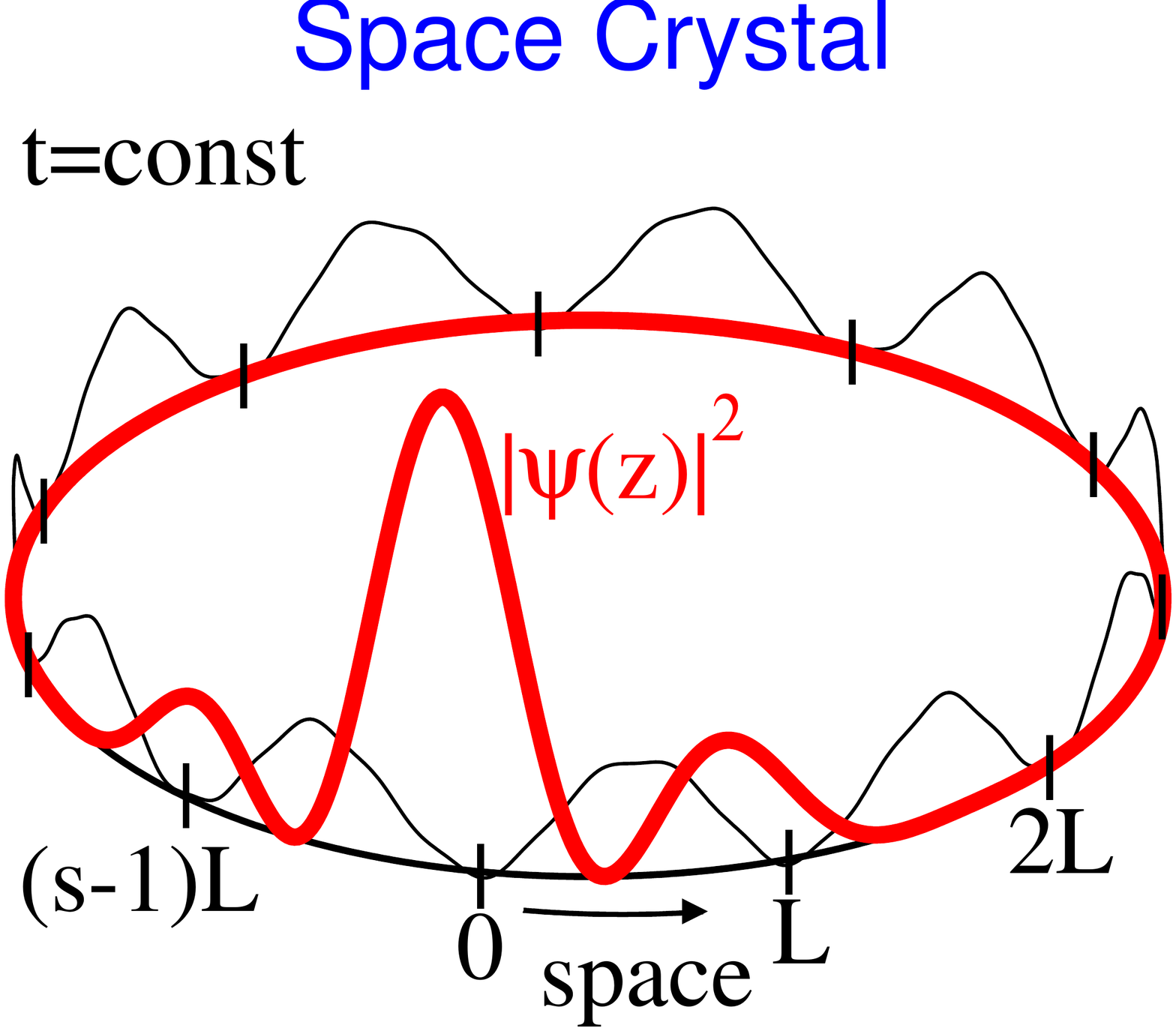}
\includegraphics[width=0.45\textwidth]{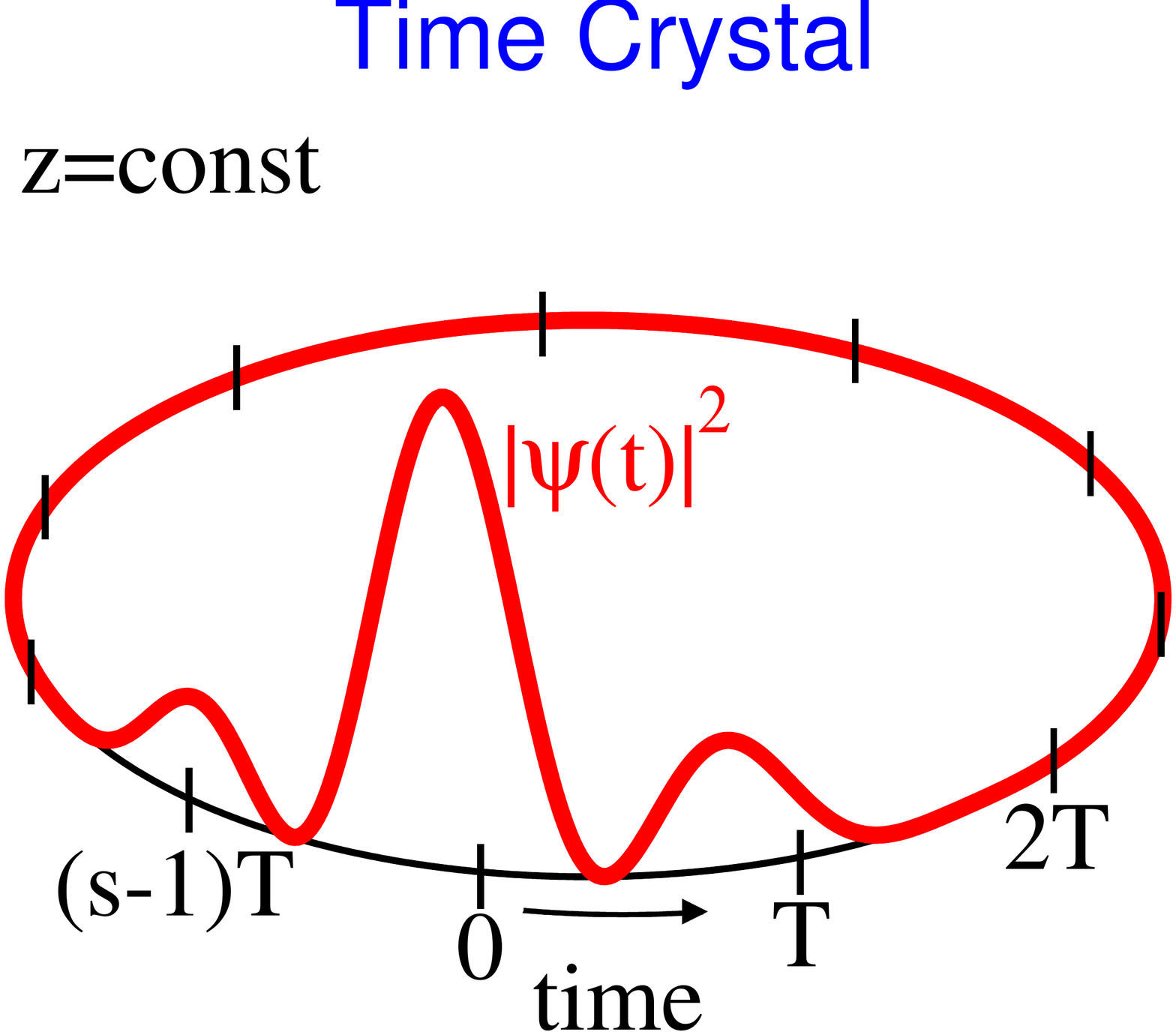}
\caption{Comparison of Anderson localization in a 1D space crystal with periodic boundary conditions (left) with Anderson localization in the time domain (right). A particle moving in the space crystal with disorder can Anderson localize, i.e., if we go around the ring, we observe an exponential localization of a particle around a certain space point. Switching from Anderson localization in space to Anderson localization in time, we have to exchange the role of space and time. We fix the position in space and ask if the probability of clicking of a detector is exponentially localized around a certain moment of time --- such behavior is repeated periodically due to the periodic boundary conditions in time. This circulating ring system is equivalent to an atom bouncing resonantly on an oscillating mirror. Reprinted from \cite{Sacha2017rev}.}
\label{singAnder}
\end{figure}

\noindent
{\it Anderson localization in time}

The standard Anderson localization phenomenon is exponential localization of eigenstates of a particle in configuration space due to the presence of a disordered potential in space \cite{Anderson1958}. Anderson localization can also be observed in momentum space (often called dynamical localization) and it is related to the quantum suppression of classical diffusion of a particle in classically chaotic systems \cite{Fishman:LocDynAnders:PRL82,MuellerDelande:Houches:2009}. It turns out that yet another version of Anderson localization is possible: Anderson localization in time due to the presence of disorder in time \cite{Sacha15a,sacha16,Giergiel2017,delande17,Giergiel2018a,Matus2021}. 

Suppose that a single particle, described by the Hamiltonian (\ref{h_init}), is driven in time like $f(t)=\lambda\cos(\omega t)+\sum_k f_k e^{ik\omega t/s}$, where the $f_k$'s are random complex numbers with $\lvert f_k\rvert \ll \lambda$. Then, the effective Hamiltonian (\ref{singheff}) reveals a similar crystalline structure as in the Hamiltonian (\ref{singheff1}) but with a weak disordered contribution \cite{Sacha15a,sacha16}. In the tight-binding approximation, Eq.~(\ref{singt-b}), there are now additional terms, $\sum_{j=1}^s \epsilon_j \lvert a_j\rvert^2$, where the $\epsilon_j$'s are real random numbers, and we arrive at the Anderson model and Anderson localization can be observed. That is, if we locate a detector close to the resonant trajectory, the probability of clicking of the detector is exponentially localized around a certain moment of time \cite{Sacha15a}. In Fig.~\ref{singAnder} we illustrate Anderson localization in the time domain and compare it with the standard Anderson localization in a space crystal. Actually, Anderson localization in time does not require an underlying crystalline structure in time. It is sufficient to drive a particle in a disordered way, $f(t)=\sum_k f_k e^{ik\omega t/s}$, where the $f_k$'s are the random numbers, and the particle can reveal localization in the time domain \cite{sacha16,Giergiel2017}. It is even possible to observe a localized-delocalized Anderson transition in the time domain if a particle moving in 3D space is properly driven \cite{delande17}. Recently, signatures of Anderson localization in time have also been analyzed in photonic time crystals, i.e., in dielectric materials where the refractive index is modulated periodically in time \cite{Sharabi2021}.
\newline

\begin{figure}[t]
\centering
\includegraphics[width=0.6\textwidth]{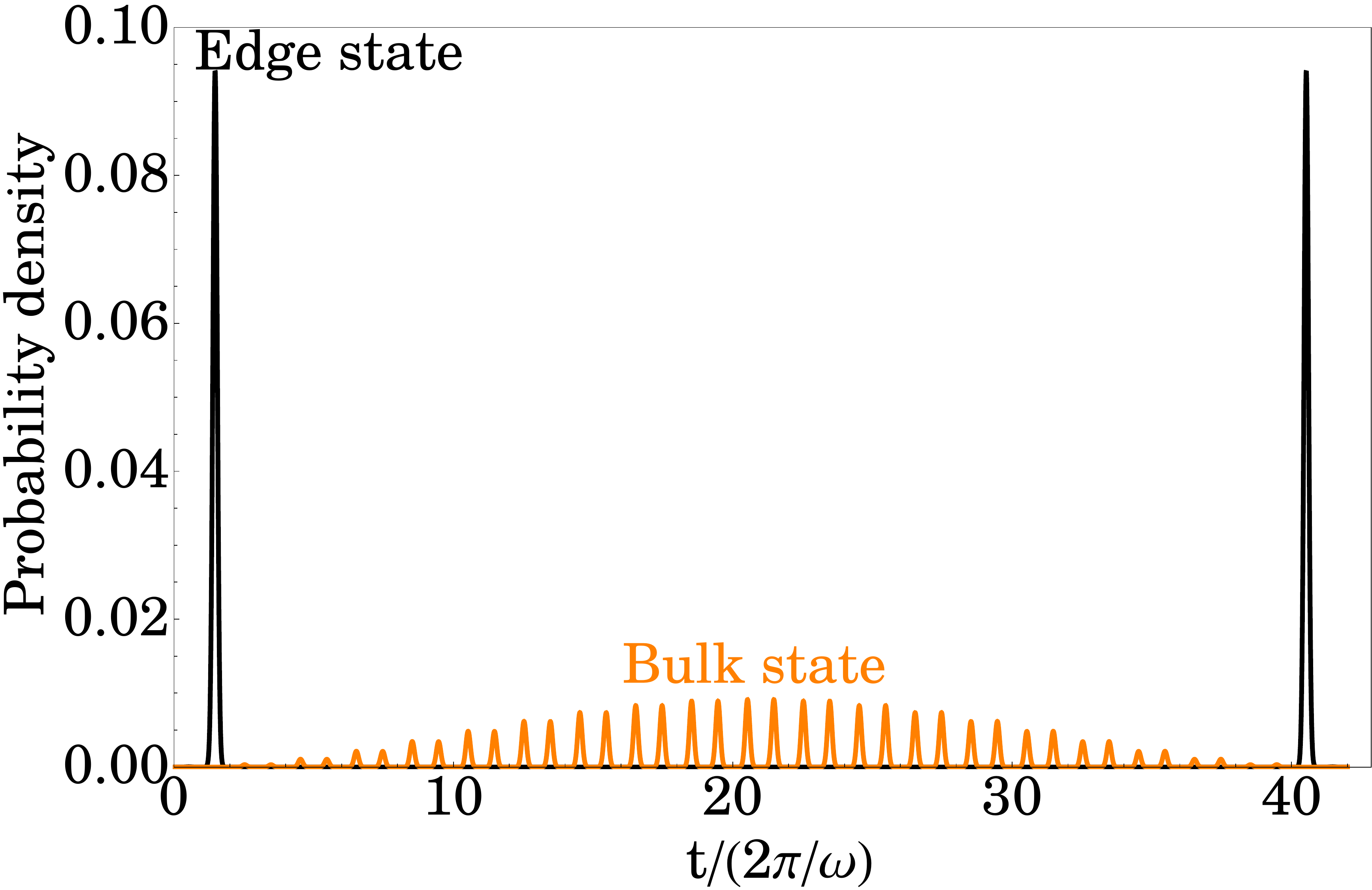}
\caption{A particle driven resonantly in time so that the effective Hamiltonian $H_{eff}$, Eq.~(\ref{singheff}), reproduces the topological Su-Schrieffer-Heeger model with an edge \cite{Su1979,Asboth2016short}. There are two kinds of eigenstates of $H_{eff}$: bulk states and topologically protected edge states. In the laboratory frame, the bulk states are delocalized along the entire resonant orbit while the edge states are localized close to the edge which appears periodically in time. The case of a $42:1$ resonance is presented in the plot. Reprinted from \cite{SachaTC2020}.}
\label{singTop}
\end{figure}

\noindent
{\it Single-particle topological time crystals}

As a second example, we present topological time crystals \cite{Giergiel2018b}. Topological insulators are solid state systems which are insulators in the bulk but possess topologically protected conducting edge or surface states which are responsible for quantization of the Hall conductance in the quantum Hall effect \cite{Hasan2010}. If a particle is resonantly driven with the help of two harmonics, i.e., $f(t)=\lambda\cos(s\omega t)+\lambda'\cos(s\omega t/2)$, then the effective potential in Eq.~(\ref{singheff}) reveals a crystalline structure with a two-point basis, $V_{eff}(\Theta)=V_0\cos(s\Theta)+V_0'\cos(s\Theta/2)$. In the quantum description of the particle, the first two energy bands of $H_{eff}$ can be described by the Su-Schriefer-Heeger (SSH) model which is a tight-binding model with staggered hopping amplitudes \cite{Su1979,Asboth2016short}. Depending on the ratio of the hopping amplitudes, the system can be characterized by a zero or nonzero winding number. In the latter case, the system is topologically nontrivial and when we introduce a narrow barrier in $V_{eff}(\Theta)$, we deal with a topological system with an edge and eigenstates localized close to it. The barrier can be realized by means of an additional modulation of the system in time, i.e., $f(t)=\lambda\cos(s\omega t)+\lambda'\cos(s\omega t/2)+\sum_kf_k e^{ik\omega t}$, where the $f_k$'s are chosen so that $V_{eff}(\Theta)$ acquires an additional contribution in the form of a narrow barrier \cite{Giergiel2018b}. In the presence of the edge, apart from bulk states which are delocalized along the entire lattice, there are also two eigenstates (with eigenenergies located in the gap of the spectrum) which are localized close to the edge. These edge states appear localized in time when we observe the particle in the laboratory frame \cite{Giergiel2018b}, see Fig.~\ref{singTop}. 

Lusting {\it et al.} have considered a topological photonic time crystal with edges, i.e., the photonic crystal was finite because it was turned on at some moment of time and later turned off. They observed an exponential increase of amplitudes of electromagnetic waves propagating in a dielectric material if the wavenumber was located in the gap \cite{Lustig2018}.
\newline

\begin{figure}[t]
\centering
\includegraphics[width=0.45\textwidth]{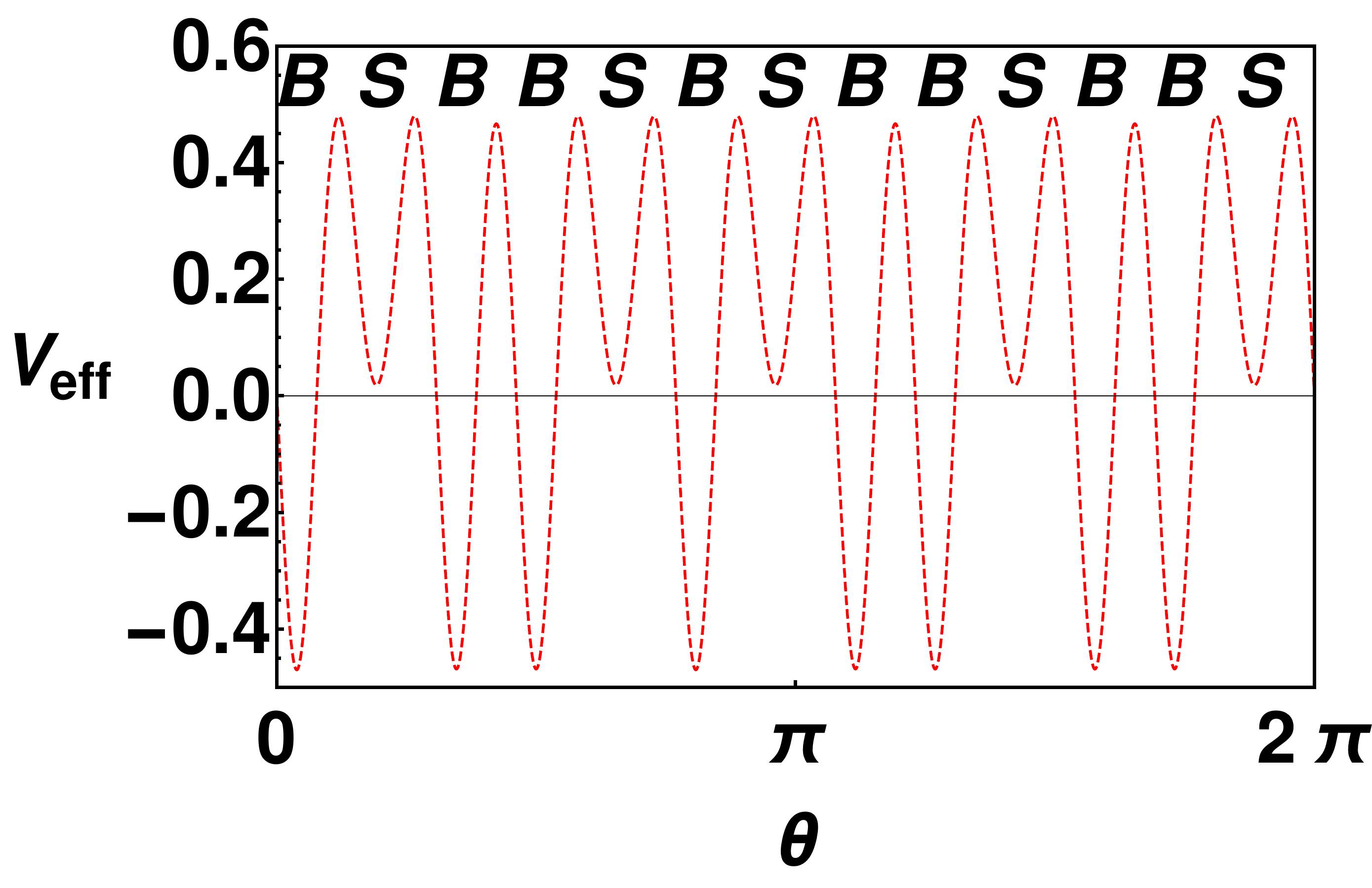}
\includegraphics[width=0.45\textwidth]{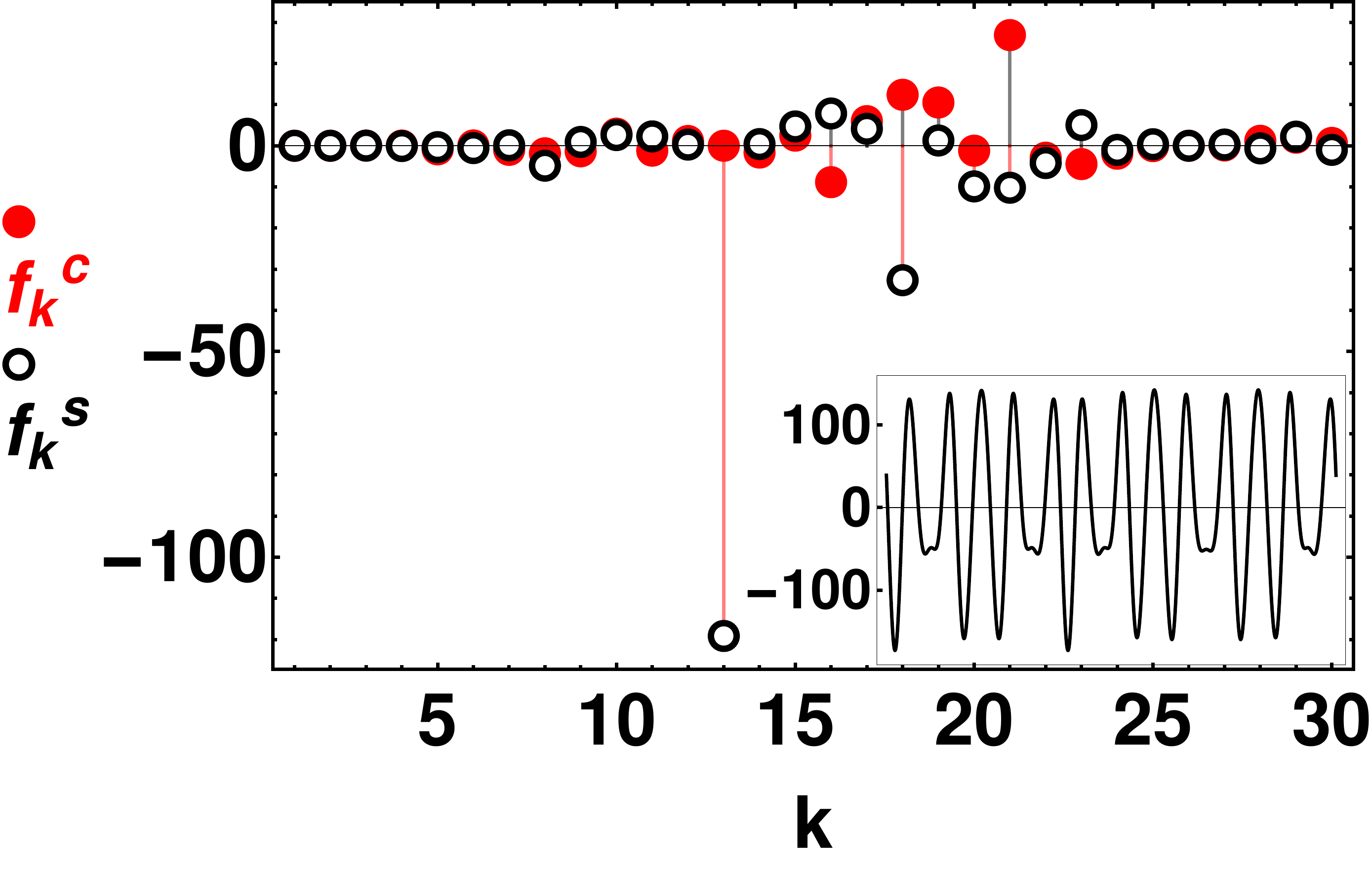}
\caption{Left panel: quasi-crystal structure in time. The motion of the mirror is chosen so that the effective potential $V_{eff}(\Theta)$ in Eq.~(\ref{singheff}) is a finite fragment of the Fibonacci sequence of small ($S$) and big ($B$) potential wells. Right panel: Fourier components of the mirror oscillations, $f(t)/\omega^2=\sum_k [f_k^c\cos(k\omega t)+f_k^s\sin(k\omega t)]$ --- the inset shows $f(t)/\omega^2$ over one period $T=2\pi/\omega$. Reprinted from \cite{Giergiel2018}.}
\label{singquasi}
\end{figure}

\noindent
{\it Time quasi-crystal structures}

We have seen that different crystalline structures can be realized in the time domain by means of resonant driving of a periodically moving particle. However, one can realize nearly any shape of the effective potential in Eq.~(\ref{singheff}) if the Fourier components of $f(t)$ are properly chosen \cite{Giergiel2018}. Indeed, any potential can be expanded in the series $V_{eff}(\Theta)=\sum_mV_me^{im\Theta}$ and choosing $f_{-m}=V_m/h_{m}(I_1)$ in the $1:1$ resonant driving of a particle allows one to realize such an effective potential in Eq.~(\ref{singheff}). Figure~\ref{singquasi} shows an example where $V_{eff}(\Theta)$ possesses a quasi-crystal structure \cite{Janot1994}, i.e., it is a series of small ($S$) and big ($B$) potential wells ordered according to the Fibonacci sequence $BSBBSBSB\dots$ and transport properties of the particle in a time quasi-crystal can be investigated \cite{Giergiel2018}.

\section{Many-Body Condensed Matter Phenomena in the Time Dimension}
\label{m-b_condmat}

We have seen that a resonantly driven single particle can behave in the time domain like an electron moving in a space crystal. One may raise the question if many-body condensed matter phenomena can be observed in the time dimension too? Having the Bose-Hubbard Hamiltonian (\ref{manybh}) that describes the resonant driving of a many-body system, the investigation of many-body solid state phenomena is quite straightforward which we show in this section.
\newline

\begin{figure}[t]
\centering
\includegraphics[width=0.42\textwidth]{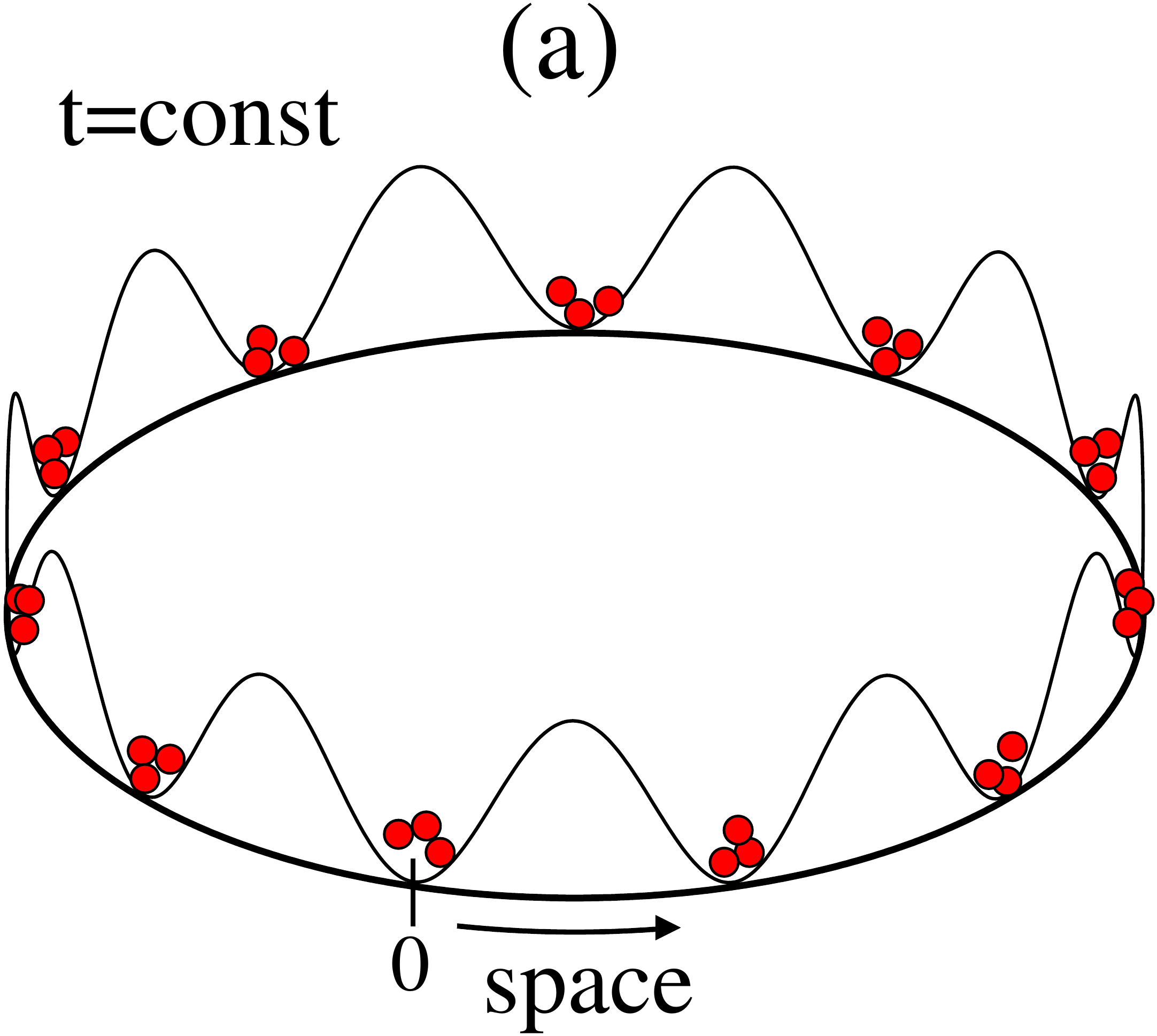}
\hfill
\includegraphics[width=0.42\textwidth]{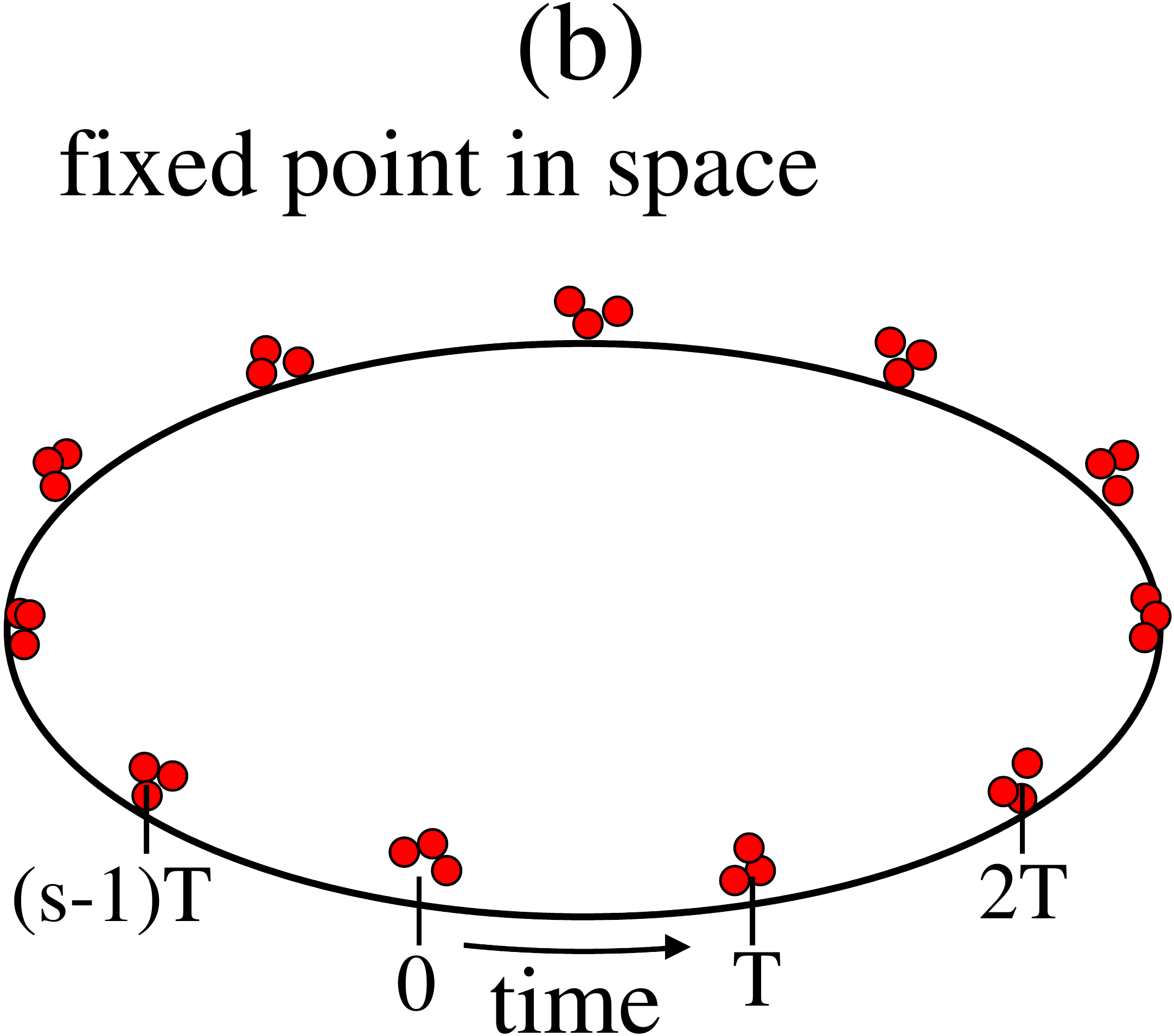}
\caption{(a) Schematic plot of the Mott insulator phase in a 1D space crystal with periodic boundary conditions. (b) Similar Mott insulator phase but in the time domain. In order to switch from space to time crystals, one has to exchange the role of space and time. That is, the position in space is fixed and we analyze how the probability of detection of particles at this fixed point changes in time. In the Mott insulator phase, a well defined number of atoms are appearing at this fixed point like on a conveyor belt or in a machine gun. 
Reprinted from \cite{Sacha15a}.}
\label{figMott}
\end{figure}

\noindent
{\it Mott insulator phases in time}

If in a resonantly driven $N$-atom system, Eq.~(\ref{manybh}), the effective repulsive on-site interactions dominate over the long-range repulsion ($U_{ii}\gg U_{ij\ne i}$) and they are sufficiently strong compared with the hopping rate ($U_{ii}N\gg J$), the system reveals a Mott-insulator phase \cite{Sacha15a}. That is, the gap between the ground and excited states of the Bose-Hubbard Hamiltonian (\ref{manybh}) is opened, fluctuations of the number of atoms in each lattice site are suppressed and the system is not compressible. This behavior is observed in the time domain. When we look at the system in the laboratory frame at a fixed position close to the resonant trajectory, we will see that the atoms arrive periodically at the observation point in well-defined portions of $N/s$ and there is no coherence between them \cite{Sacha15a}, see Fig.~\ref{figMott}. This is in contrast to the superfluid phase where the repulsive interactions are weak and in a finite lattice ($s<+\infty$) the ground state of the Bose-Hubbard model (\ref{manybh}) is a Bose-Einstein condensate. The superfluid-Mott insulator transition can be realized in the time lattice described here by changing the hopping rate $J$ which can be done by changing the amplitude of the periodic driving of the system \cite{Sacha15a}.
\newline

\noindent
{\it Many-body localization in time}

If a resonantly driven many-body system experiences temporal disorder, the many-body localization phenomenon in a time crystalline structure can be investigated \cite{Mierzejewski2017}. Indeed, when the driving of the system possesses a fluctuating contribution, there are additional terms in the effective Bose-Hubbard model (\ref{manybh}), i.e., $\sum_{j=1}^s \epsilon_j\hat a_j^\dagger \hat a_j$, where the $\epsilon_j$'s are random numbers (see Anderson localization in time described in the previous section), and the system can reveal many-body localization if the disorder is sufficiently strong. Many-body localization is characterized by a lack of dc transport, extremely slow dynamics of various correlation functions and a logarithmic growth of the entanglement entropy, and it has been extensively investigated in many-body systems in the presence of time-independent disorder \cite{pal10,schreiber15}. Here, we see that many-body localization can also be induced by temporal disorder \cite{Mierzejewski2017,SachaTC2020}.
\newline

\begin{figure}[t]
\centering
\includegraphics[width=0.49\textwidth]{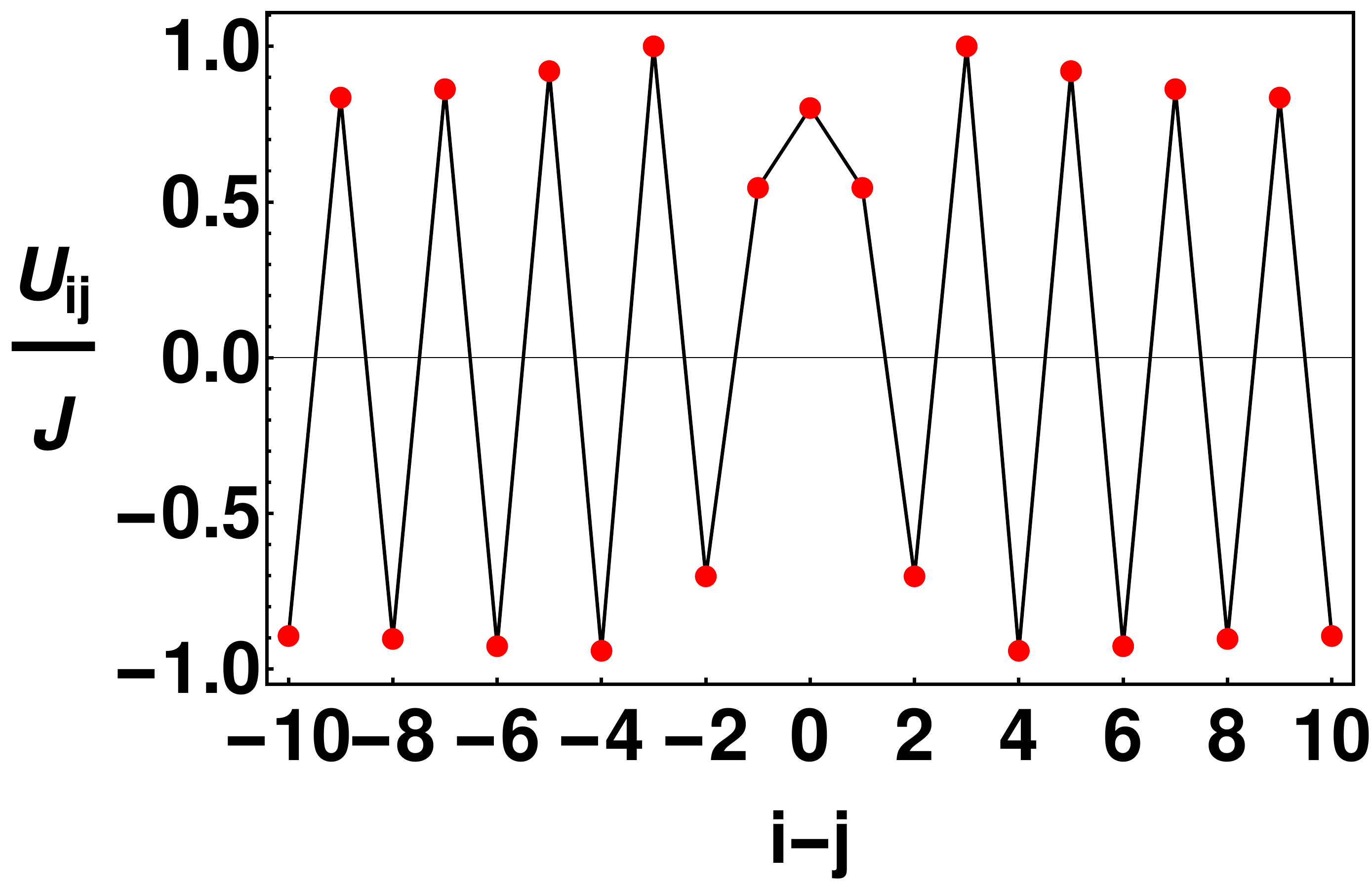}
\hfill
\includegraphics[width=0.49\textwidth]{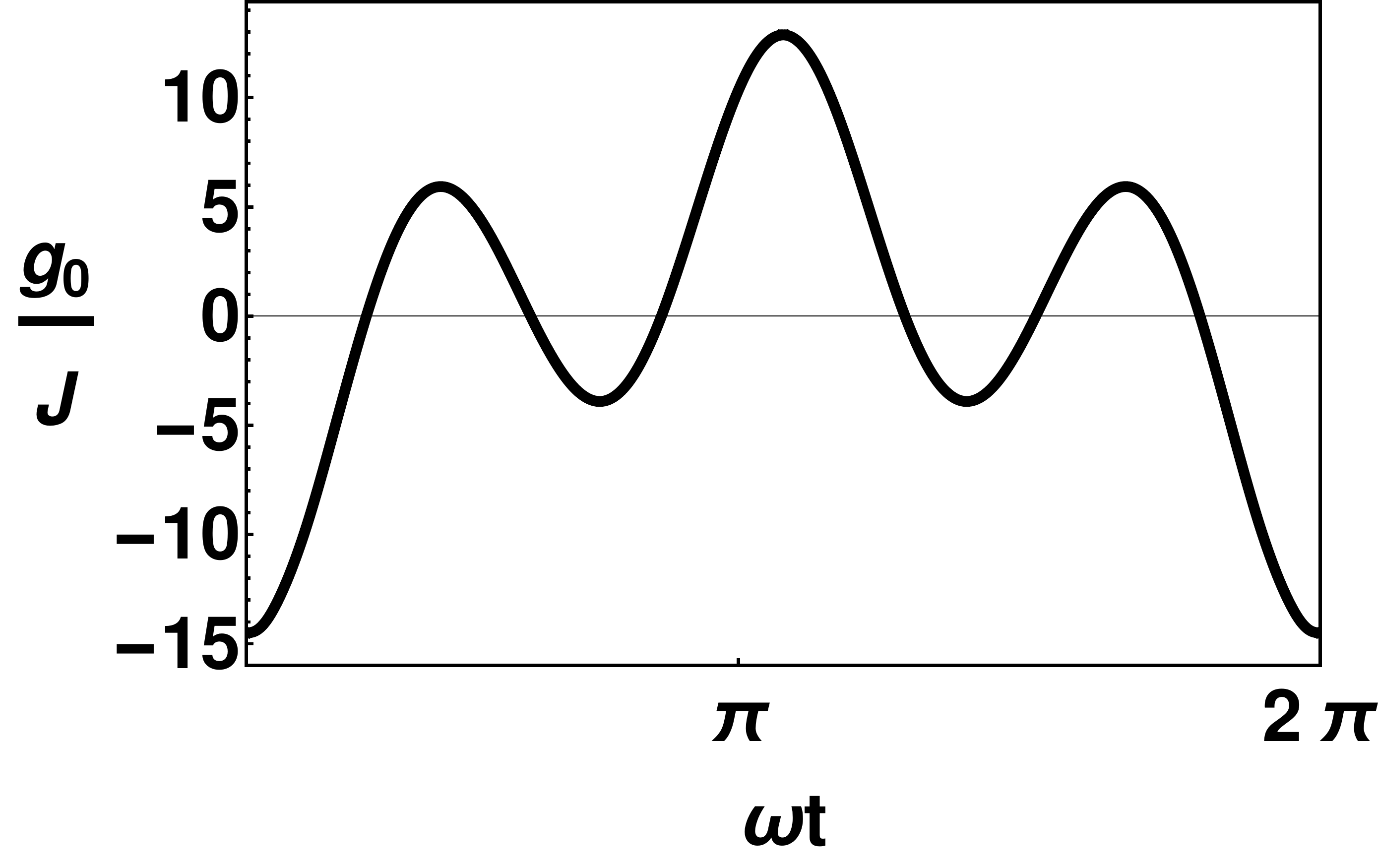}
\caption{Left panel: effective long-range interaction coefficients as a function of the distance $i-j$ between temporal lattice sites in the Bose-Hubbard Hamiltonian (\ref{manybh}) in units of the tunneling amplitude $J$. The character of the interactions changes between repulsive ($U_{ij}>0$) and attractive ($U_{ij}<0$) in an oscillatory way vs. $i-j$. Right panel: periodic changes of the original contact interaction strength $g_0(t)$ between ultracold atoms that results in the effective interactions presented in the left panel, cf. Eq.~(\ref{manyuij}).
Reprinted from \cite{Giergiel2018}.}
\label{fig_long}
\end{figure}

\noindent
{\it Long-range exotic interactions in time}

In a typical situation, the effective long-range interactions in the Bose-Hubbard model (\ref{manybh}) are negligible compared to the on-site interactions. However, if we modulate periodically in time the contact interaction strength $g_0(t)$ in Eq.~(\ref{manyuij}), we can make the long-range interactions significant and even engineer how they change with the distance $\rvert i-j\rvert$ between lattice sites \cite{Giergiel2018}. In ultracold atoms experiments $g_0$ can be modulated in time by modulating the external magnetic field around a Feshbach resonance \cite{Chin2010}. If $g_0(t)$ changes in time with the period $sT$, the interaction coefficients can be written as $U_{ij}=\int_0^{sT}dt \;u_{ij}(t) g_0(t)$ where $u_{ij}(t)=(2-\delta_{ij})\int dx \rvert w_i\rvert^2\rvert w_j\rvert^2/sT$. The latter can be considered as a matrix with $(i,j)$ and $t$ treated as indices of rows and columns, respectively. The left singular vectors of the matrix $u_{ij}(t)$ determine which sets of the interaction coefficients $U_{ij}$ can be realized in a system while the right singular vectors tell us how we should modulate $g_0(t)$ in order to achieve them. Such an approach allows one to engineer very exotic long-range interactions which do not exist in nature --- for example, interactions that change the repulsive and attractive character in an oscillatory way with an increase of $\rvert i-j\rvert$, see Fig.~\ref{fig_long} \cite{Giergiel2018}. 

In the case of many-body phase space crystals investigated by Guo and coworkers \cite{Guo2016,Liang2017,Guo2020}, it is also possible to realize effective long-range interactions between ultracold atoms but they are long-range in phase space.
\newline

\noindent
{\it Many-body topological time crystals}

If $g_0$ can be modulated not only in time but also in space, there are additional possibilities to engineer different effective long-range interactions. For example, if $g_0(z,t)=\sum_{m=0}^2\alpha_m(t)z^m$, the coefficients $\alpha_m(t)$ can be chosen so that the Bose-Hubbard Hamiltonian (\ref{manybh}) reduces to \cite{Giergiel2018b} 
 \be
\hat H_F\approx -\frac{J}{2}\sum_{j=1}^s\left(\hat a_{j+1}^\dagger\hat a_j+H.c.\right)+\frac{U}{2} \sum_{j=1}^s\hat n_j(\hat n_j-1)+ V\sum_{j=1}^s\hat n_{j+1}\hat n_j,
\label{manyhaldane}
\ee
where $\hat n_j=\hat a_j^\dagger\hat a_j$. This system can reveal the bosonic analog of the topological Haldane insulator in a spin-1 chain \cite{Haldane83,Haldane83a,Torre2006,
Rossini2012}. That is, between the Mott insulator phase, which is observed if the repulsive on-site interactions are dominant, and the density-wave phase, observed for dominant repulsive nearest-neighbor interactions, there is the topological Haldane insulator phase related to a highly nonlocal
string order parameter \cite{Kennedy92,Torre2006}. All of these phases can be realized in ultracold atoms bouncing resonantly on an oscillating atom mirror in the presence of an external Feshbach magnetic field that changes in time and space in a proper way \cite{Giergiel2018b}.

\section{Higher-dimensional time lattices}
\label{md_timecrystals}

Although time is a single degree of freedom, crystalline structures in
time with the properties of 2D (or 3D) condensed matter systems can be created by a BEC bouncing resonantly between two (or three)
oscillating atom mirrors \cite{SachaTC2020,Giergiel2018,Zlabys2021,Kuros2021,Giergiel2021}.

\begin{figure}[t]
\centering
\includegraphics[width=0.6\textwidth]{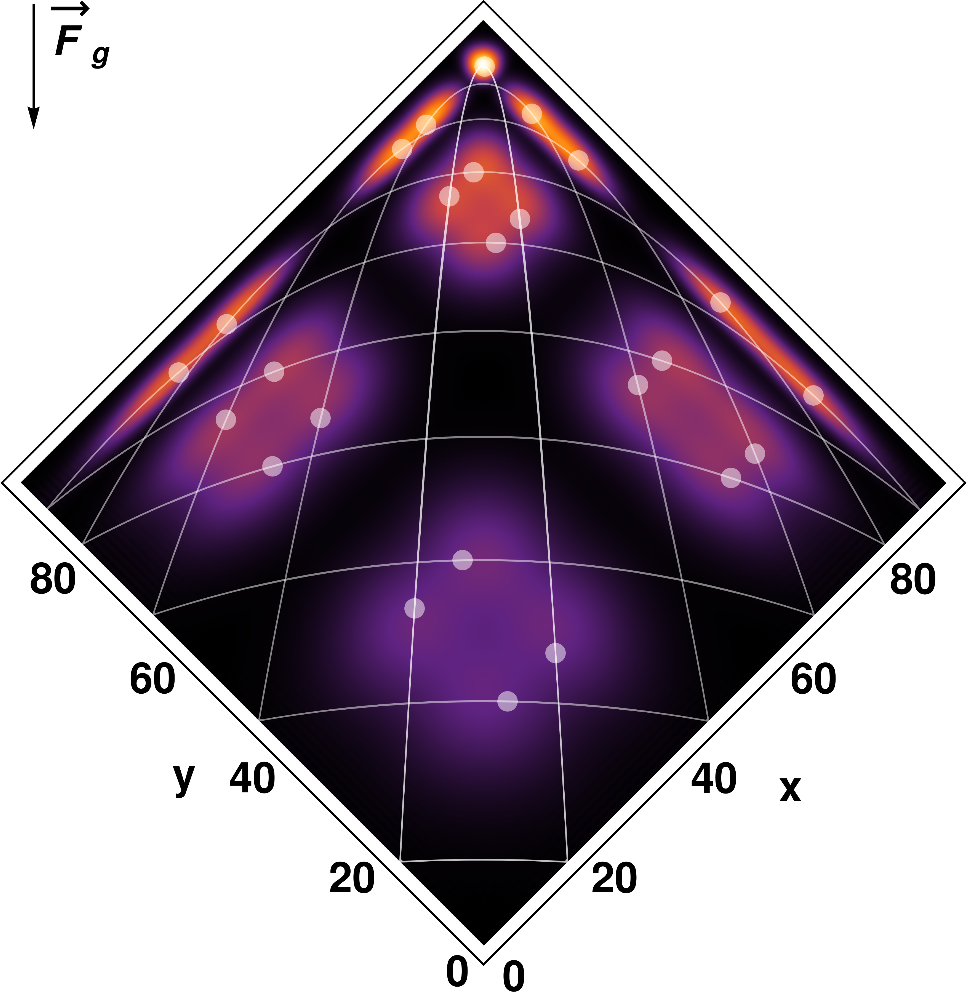}
\caption{Trajectories of atoms bouncing between two oscillating mirrors, located at $x= 0$ and $y = 0$, on a two-dimensional $5\times 5$ time lattice. Reprinted from \cite{Giergiel2018}.}
\label{time_latice}
\end{figure}

We consider the case of atoms bouncing resonantly between two
orthogonal mirrors, located at $x= 0$ and $y = 0$,
oscillating at the same frequency $\omega$ and inclined at 45\textsuperscript{o} to the gravitational force (Fig.~\ref{time_latice}), but this can be extended to three orthogonal oscillating mirrors. In this case
the motion of a single particle separates into two \emph{independent}
one-dimensional motions along the directions orthogonal to the mirrors.
Averaging the 2D equivalent of the single-particle Hamiltonian over time
leads to the 2D classical secular effective Hamiltonian, cf. Eq.~\eqref{singheff1}, \cite{SachaTC2020} 
\be
H_{\text{eff}} = \ \frac{P_{+}^{2} + P_{-}^{2}}{2{\rvert m}_{\text{eff}}\rvert} + \ V_{0}\lbrack\cos\left( s\Theta_{x} \right) + \cos{\left( s\Theta_{y} \right)\rbrack}.
\label{heff2D}
\ee

Switching to $N$-boson particles and restricting to the resonant
Hilbert subspace, one obtains a 2D Bose-Hubbard model similar to the 1D
model, Eq.~\eqref{manybh}. Restricting to the first energy band in the quantized
version of Eq.~\eqref{heff2D} leads to the tight-binding approximation. In the Hilbert
subspace corresponding to the first energy band we can define a basis of
the Wannier states
$W_{\mathbf{i}}\left( \Theta_{x}{,\Theta}_{y} \right)$ localized
in minima of the potential \(V_{0}\) in Eq.~\eqref{heff2D}. In the laboratory frame,
these 2D Wannier states $W_{\mathbf{i} = i_{x}{, i}_{y}}\left(x,y,t\right)$
are products of the 1D localized wave-packets
$w_{i_{x}}\left(x,t\right)$ and
$w_{i_{y}}\left(y,t\right)$ moving along the classical
orbits.

When the bouncing atoms fulfill the $s:1$ resonance criterion
with each mirror, a square $s\times s$ time lattice is created
(see Fig.~\ref{time_latice} for the case of a $5\times 5$ time lattice). In the
Bose-Hubbard model of a 2D time lattice, the on-site interactions $U_{\mathbf{ii}}$ typically dominate over the long-range
inter-site interactions  $U_{\mathbf{ij\ne i}}$ and for sufficiently
strong repulsive interactions, the ground state of the system
corresponds to a Mott insulating phase, which for unit filling of the
lattice is a Fock state in which a single bosonic atom occupies each localized
wave-packet.

Use of a 2D time lattice allows a range of 2D condensed matter phenomena
to be investigated with two oscillating mirrors. For example,
it is possible to realize spontaneous formation of time quasi-crystals
when the ratio of the resonance numbers along the $x$ and $y$
directions, $s_{x}/s_{y}$, approximates the golden number
$(1 + \sqrt{5})/2 \approx 1.618$ \cite{Giergiel2018c}. Then, when the discrete time
translation symmetry (defined by the oscillation period of the mirrors)
is spontaneously broken, the sequence of atoms reflected from one mirror
and the other mirror forms a Fibonacci quasi-crystal in time \cite{SachaTC2020,Giergiel2018c}.

In a 2D time lattice, one can also realize gradual breaking of the
discrete time-translation symmetry when the motion along one of the two
orthoganol directions breaks the symmetry and with a range of system
parameters the motion along the other direction also reveals symmetry
breaking \cite{Kuros2021}.
\newline

\noindent
\emph{M\"obius strip geometry}

We now consider the case when the two mirrors are oriented at
45\textsuperscript{o} to each other to form a wedge with one mirror
located in the vertical direction $x - y = 0$ and the other located at $x = 0$ \cite{Giergiel2021}. When a particle strikes the vertical mirror, its momenta are
exchanged $p_{x}\leftrightarrow p_{y}$, whereas when it strikes
the mirror located at $x = 0$, the momentum $p_y$ remains
the same while the momentum $p_{x} \rightarrow - p_{x}$. In the
following we suppose the mirror located at \(x = 0\) oscillates with the
driving function $f_{x}(t) =-(\lambda_1/\omega^2)\cos(\omega t)-(\lambda_2/4\omega^2)\cos(2\omega t)$
while the vertical mirror oscillates with
$f_{x - y}\left( t \right) = -(\lambda_3/4\omega^2)\cos(2\omega t+\phi)$,
where $\lambda_{1,2,3}$ are the oscillation amplitudes and $\phi$ is a constant phase.

Transforming to the action-angle variables $I_{\pm} =I_{y}\pm I_{x}$ and
$\theta_{\pm} = (\theta_{y}\pm\theta_{x})/2$ and
switching to the frame oscillating with the mirrors,
$\Theta_{+} = \theta_{+} - \omega t/s$ and
$\Theta_{-} = \theta_{-}$, leads to the classical secular effective
Hamiltonian \cite{Giergiel2021} 
\bea
H_{eff} &=& -\frac{P_{+}^{2} + P_{-}^{2}}{2{\vert m}_{eff}\vert} - \frac{\lambda_{2}}{2\omega^{2}}\cos(2s\Theta_{+})\cos(2s\Theta_{-}) - \frac{{2\lambda}_{1}}{\omega^{2}}\cos(s\Theta_{+})\cos(s\Theta_{-}) 
\cr &&
+ \frac{\lambda_{3}}{4\omega^{2}}\cos(2s\Theta_{+}+\phi),
\label{heff_lieb}
\eea
where $P_{\pm} =I_{\pm} - I_{\pm}^{0}$, 
$\vert m_{eff}\vert = \left(3I_{\pm}^{0} \right)^{4/3}/\left( 2\pi^{2} \right)^{1/3}$
and $I_{\pm}^{0}$ is the resonant value of the action $I_{\pm}$. The
Hamiltonian \eqref{heff_lieb} describes a particle with negative effective mass $- {\vert m}_{eff}\vert$ in the presence of an
\emph{inseparable} lattice potential which is moving on a M\"obius strip
because at $\Theta_{+} = \pi$, $\Theta_{\pm}$ flips
to $\pi - \Theta_{\pm}$.

\begin{figure}[t]
\centering
\includegraphics[width=0.9\textwidth]{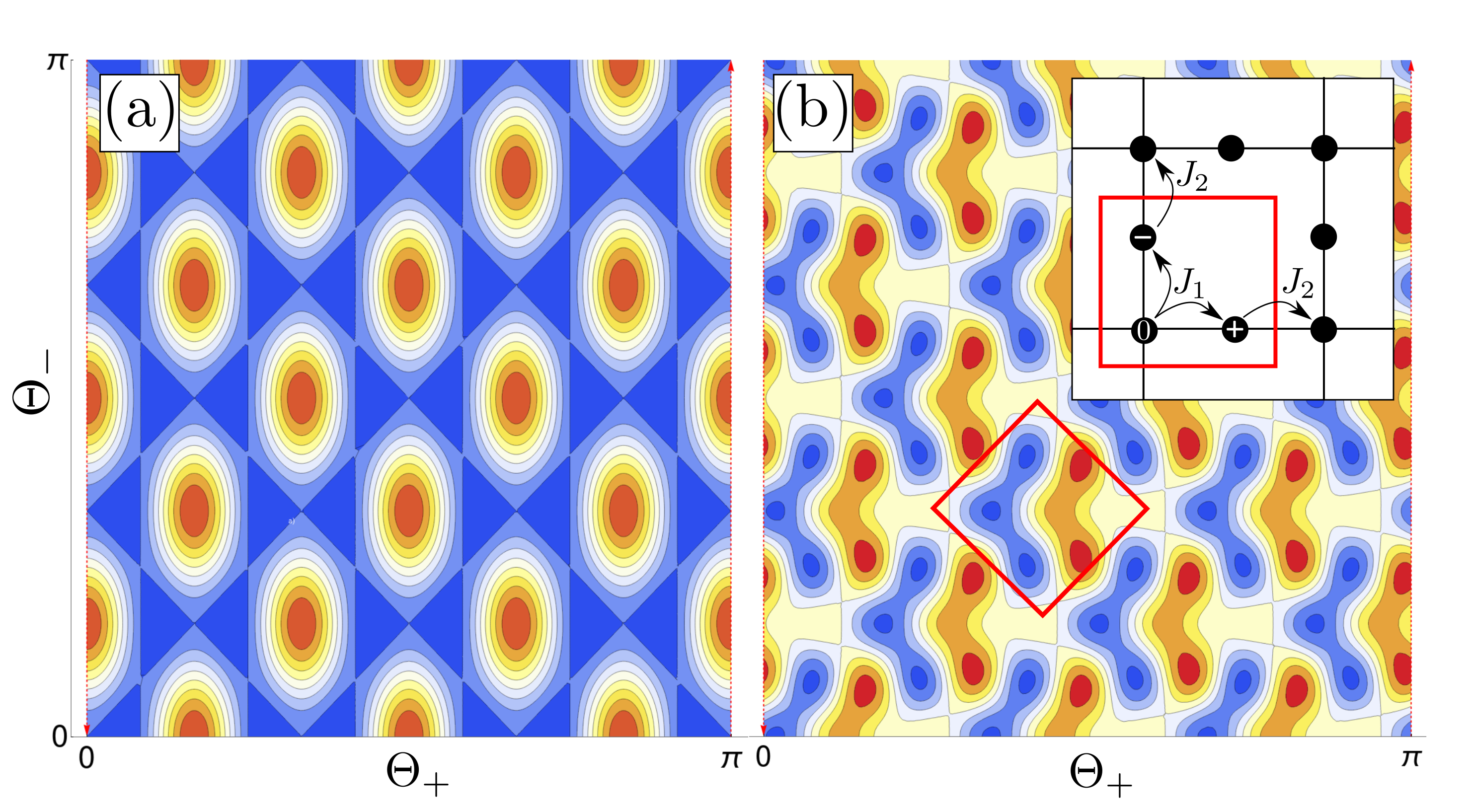}
\caption{(a) Honeycomb and (b) Lieb time lattices presented in the
frame moving along a resonant trajectory. Atoms loaded to lattices
fulfill the boundary conditions of the M\"obius strip geometry,
i.e., points $\{\Theta_{+} =\pi,\Theta_{-}\}$ are identified
with $\{\Theta_{+} = 0,\pi - \Theta_{-}\}$. Dark blue regions correspond to the lowest energies of a particle with negative effective mass. Reprinted from \cite{Giergiel2021}.}
\label{lieb_latice}
\end{figure}

Use of different parameters of the mirrors' oscillations allows one to
realize different crystalline structures of the effective potential in \eqref{heff_lieb}. For example,
for $\lambda_{3}/\lambda_{1} = 4$, $\lambda_{2} = 0$
and $\phi = 0$, a honeycomb time lattice can be realized (Fig.~\ref{lieb_latice}(a)),
while for $\lambda_{3}/\lambda_{1} = 4$,
$\lambda_{3}/\lambda_{2} = 1.62$
and $\phi =\pi/4$, a Lieb `decorated' square time lattice
with a flat band -- in which the dynamics of the atoms is governed solely
by interactions -- is created (Fig.~\ref{lieb_latice}(b)) \cite{Giergiel2021}.
\newline

\noindent 
{\it Six-dimensional time-space crystals}

We live in a 3D space and spatial crystals of dimension up to three can be observed and investigated. In the previous sections we have seen that {\it time} appears as a degree of freedom where a crystalline behavior can be created too. Combining space and time lattices, we should be able to create crystals of dimension higher than three. Surprisingly, such time-space crystals can be even six-dimensional as we will show in the following \cite{Zlabys2021}. 

Suppose that ultracold atoms are prepared in a 1D optical lattice potential created by means of an optical standing wave \cite{Pethick2002}. Suppose also that the optical lattice is periodically shaken. If the atoms do not interact, the description of the system can be reduced to the following {\it single-particle} Hamiltonian
\be
H=\frac{p_x^2}{2}+V_0\sin^2(x-\lambda\cos\omega t),
\label{time-space1}
\ee
where $V_0$ is the amplitude of the optical potential while $\lambda$ and $\omega$ denote the amplitude and frequency of the shaking of the potential. We assume that the depth of the optical lattice potential is sufficiently large that the atoms can perform periodic motion in the lattice wells which is resonant with the shaking. If the $s:1$ resonance condition is fulfilled, $s$ localized wave-packets can form in each well which evolve along the resonant trajectories with the period $s2\pi/\omega$. Restricting to the resonant Hilbert subspace we obtain an effective tight-binding Hamiltonian \cite{Zlabys2021}
\be
H_{eff}\approx -\frac12 \sum_{i,\alpha,j,\beta}J_{i,\alpha}^{j,\beta}\;a^*_{j,\beta}\;a_{i,\alpha},
\label{time-space_heff}
\ee
which describes a 2D time-space lattice, where $(i, j)$ and $(\alpha,\beta)$ denote indices of spatial and temporal lattice sites, respectively. To obtain Eq.~\eqref{time-space_heff} we have assumed that the atom is described by the wavefunction $\psi(x,t)=\sum_{i,\alpha}a_{i,\alpha}w_{i,\alpha}(x,t)$, where $w_{i,\alpha}(x,t)$ is a localized wave-packet evolving periodically along the resonant trajectory in the $i$-th well of the optical lattice potential. In each well there are $s$ such wave-packets, i.e., $\alpha=1,\dots,s$. The parameters $J_{i,\alpha}^{j,\beta}$ describe hopping of the atom between the spatial sites $i$ and $j$ and between the temporal sites $\alpha$ and $\beta$.

The 1D optical lattice problem can be easily generalized to a 3D optical lattice potential which is periodically shaken along the three independent spatial directions 
\bea
H&=&\frac{p_x^2+p_y^2+p_z^2}{2}+V_0[\sin^2(x-\lambda\cos\omega t)+\sin^2(y-\lambda\cos\omega t)
\cr
&&
+\sin^2(z-\lambda\cos\omega t)].
\label{time-space2}
\eea
If the $s:1$ resonance condition is fulfilled for the motion of the atom along each spatial direction, then we can create 3D localized wave-packets which are products of the wavefunctions of the 1D wave-packets, i.e.,
\be
W_{\vec i,\vec \alpha}(\vec r,t)=w_{i_x,\alpha_x}(x,t)
\;w_{i_y,\alpha_y}(y,t)
\;w_{i_z,\alpha_z}(z,t).
\ee
Within each well of the 3D optical lattice potential there is a $s\times s\times s$ time lattice whose temporal sites are labeled by three indices $\alpha_x$, $\alpha_y$ and $\alpha_z$. The wells of the optical lattice potential are labeled by three spatial indices $i_x$, $i_y$ and $i_z$. Thus, all together the wave-packets $W_{\vec i,\vec \alpha}(\vec r,t)$ are labeled by six indices corresponding to a 6D time-space lattice, and 6D condensed matter phenomena can be investigated. In Ref.~\cite{Zlabys2021} it was shown how to realize an artificial gauge potential by tilting the 6D time-space lattice along the temporal directions and employing so-called photon-assisted tunneling. This allowed the realization of the 6D quantum Hall effect. 

It is worth mentioning that we consider here deep optical lattice potentials which can host many energy bands, and highly excited bands (with energies close to the barrier between the potential wells) are resonantly coupled by the shaking. This is in contrast to a more typical situation when an optical lattice potential is shallow and resonant coupling of the ground and first excited energy bands are investigated \cite{Sowinski2012,Lacki2013,Li2016}.

\section{Conclusions}

Big discrete time crystals created by a Bose-Einstein condensate of ultracold atoms bouncing resonantly on an oscillating atom mirror provide a highly flexible platform for investigating a broad range of condensed matter phenomena in the time dimension. By choosing suitable Fourier components in the periodic drive of the atom mirror, such a system allows us to construct effective temporal lattice potentials of almost any shape and to readily vary the geometry of the time lattice. By modulating the interparticle scattering length via a magnetically tuneable Feshbach resonance, such a system also allows us to precisely control the effective interparticle interaction and to engineer exotic long range interactions in the time lattice. Use of a bouncing BEC system also allows us to construct higher-dimensional time lattices, involving the bouncing of a BEC between two (or three) orthogonal oscillating mirrors and also between two oscillating mirrors oriented at 45-degrees. The latter configuration supports a versatile Möbius strip geometry that can host a variety of time lattices such as a honeycomb time lattice and a Lieb square time lattice. By combining 3D time lattices with 3D space lattices, it is possible to create 6D time-space lattices.

Condensed matter phenomena in the time dimension that have been predicted to date include Anderson and many-body localization due to temporal disorder, topological time crystals, quasi-crystal structures in time, and Mott insulator phases in time. The application of big discrete time crystals to condensed matter physics in the time dimension has opened up a rich new area of research ready to be exploited experimentally. 

\section{Statements and Declarations}

\noindent
{\bf Ethical approval and consent to participants} 

\noindent
The authors declare they have upheld the integrity of the scientific record.
\newline 

\noindent
{\bf Consent for publication}  

\noindent
The authors give their consent for publication of this article.
\newline 

\noindent
{\bf Availability of data and materials}

\noindent
The data generated during the current study are available from the contributing author upon reasonable request.
\newline 

\noindent
{\bf Competing interests} 

\noindent
The authors have no competing interests to declare that are relevant to the content of this article.
\newline 

\noindent
{\bf Funding}

\noindent
This work was supported by the Australian Research Council Discovery Grant No. DP100100815 and the National Science Centre, Poland via Project No.~218/31/B/ST2/00319.
\newline 

\noindent
{\bf Authors’ contributions} 

\noindent
PH and KS equally contributed to all aspects of the manuscript. Both authors read and approved the final manuscript.
\newline 

\noindent
{\bf Acknowledgements} 

\noindent
The authors thank K. Giergiel, E. Anisimovas, B. Dalton, D. Delande, Chu-Hui Fan, W. Golletz, C. Gunawardana, A. Kosior, A. Kuro\'s, M. Lewenstein, M. Mierzejewski, A. Sidorov, A. Singh, A. Syrwid, S. Tojo, T. Tran, J. Wang, A. Zaheer, J. Zakrzewski and G. \v{Z}labys for fruitful discussions.
\newline 

\noindent
{\bf Authors’ information} 

\noindent
Peter Hannaford 

\noindent
Optical Sciences Centre  

\noindent
Swinburne University of Technology 

\noindent
Hawthorn  

\noindent
Victoria 3122, Australia 

\noindent
phannaford@swin.edu.au 

\noindent
ORCID Number: 0000-0001-9896-7284 
\newline

\noindent
Krzysztof Sacha 

\noindent
Institute of Theoretical Physics 

\noindent
Jagiellonian University 

\noindent
ulica Profesora Stanislawa Lojasiewicza 11  

\noindent
PL-30-348 Krakow, Poland 

\noindent
krzysztof.sacha@uj.edu.pl 

\noindent
ORCID Number: 0000-0001-6463-0659



\end{document}